\pdfoutput=1

\documentclass[a4paper,11pt,reqno,final]{amsart}

\usepackage{epsfig}
\usepackage{psfrag}

\graphicspath{{./dessins/}}

\usepackage{amsmath}
\usepackage{amsfonts}
\usepackage{amsthm}

\usepackage[T1]{fontenc}

\usepackage[latin1]{inputenc}

\usepackage[english,french]{babel}

\usepackage[centerlast,small]{subfigure}

\usepackage[style=alphabetic,maxbibnames=10,isbn=false,url=false,abbreviate=false,giveninits=true,backend=biber]{biblatex}
\newtheorem{theorem}{Théorème}[section]

\theoremstyle{definition}

\theoremstyle{remark}
\newtheorem{remark}[theorem]{Remarque}

\numberwithin{equation}{section}

\newcommand{\Er}{\mathbb{R}}
\newcommand{\Erp}{\mathbb{R_{\, +}}}

\newcommand{\En}{\mathbb{N}}

\makeatletter
\renewcommand{\@seccntformat}[1]{\large{\csname the#1\endcsname}.
\hspace{0.5em}}
\renewcommand{\section}{\@startsection {section}{1}{0mm}%
                                   {-\baselineskip}%
                                   {0.5\baselineskip}%
                                   {\sffamily\large\upshape\bfseries}}
\renewcommand{\subsection}{\@startsection {subsection}{2}{0mm}%
                                   {-0.5\baselineskip}%
                                   {0.5\baselineskip}%
                                   {\sffamily\normalsize\upshape
                                   \bfseries}}
\makeatother

\bibliography{morpho_JBPL_2019_arXivX5}

\begin{document}
\selectlanguage{french}

%%%%%%%%%%%%%%%%%%%%%%%%%%%%
% TITLE (for amsart class)
%%%%%%%%%%%%%%%%%%%%%%%%%%%%

%%%%%%%%% Begin of title

\title%
[Caractérisation morphométrique d'individus]%
{Modèle géométrique pour une Caractérisation morphométrique d'individus dans des conditions \textit{in situ}}

\author{Jérôme BASTIEN}
\author{Pierre LEGRENEUR}

\date{\today}

\address{%
Laboratoire Inter-universitaire de Biologie de la Motricité\\
   POLYTECH\\
   Université Claude Bernard - Lyon 1\\
   15 Boulevard André LATARJET\\
   69622 Villeurbanne Cedex\\
France
}

\address{Laboratoire Inter-universitaire de Biologie de la Motricité\\
   U.F.R.S.T.A.P.S.\\
   Université Claude Bernard - Lyon 1\\
   27-29, Bd du 11 Novembre 1918\\
   69622 Villeurbanne Cedex\\
France}

\email{jerome.bastien@univ-lyon1.fr}

\email{pierre.legreneur@univ-lyon1.fr}

\begin{abstract}
Notre objectif a  été de développer une méthode non invasive d'évaluation des caractéristiques morphométriques des individus. Après modélisation de l'individu en une chaîne poly-articulée de segments rigides et indéformables, ceux-ci sont modélisés géométriquement sous la forme de volumes. Pour ce faire, chaque segment est photographié selon 2 plans orthogonaux et chaque image est digitalisée selon un nombre quelconque de points. En supposant que la frontière de chacun des membres peut être décrite en coordonnées cylindriques, la frontière est reconstruite par interpolation. Les coordonnées des points des différents membres sont obtenues par coupe successives, à cote croissante, ce qui permet de ramener le problème à une suite d'interpolations en coordonnées polaires. Enfin, la masse volumique  de chaque segment étant supposée connue et homogène, les principales données morphométriques sont calculées grâce aux coordonnées polaires.
\end{abstract}

%%%%%%%%% End of title

\maketitle

%%%%%%%%%%%%%%%%%%%%%%%%%%%%%%%%%%%%%%%%%%%%%%%%%%%%%%%%%%%%%
%%%%%%%%%%%%%%%%%%%%%%%%%%%%%%%%%%%%%%%%%%%%%%%%%%%%%%%%%%%%%
\section{Introduction}
\label{introduction}

L'une des problématiques essentielles en biomécanique humaine ou animale est de déterminer les caractéristiques morphométriques des individus en mouvement. Ces caractéristiques sont la position des centres de masse des segments de l'individu modélisé sous la forme d'un système poly-articulé, les masses de ces segments et leurs matrices d'inertie. La connaissance de ces données est essentielle dans le sens où elle nous permettent de calculer à chaque instant la position du centre de gravité de l'individu ainsi que par dynamique inverse ou dynamique directe les forces et moments interarticulaires au cours du mouvement.
Il existe dans la littérature des tables anthropométriques
\cite{Zatsiorsky1983,winter2009}
ou 
morphométriques
\cite{Wells1987,Legreneur2012}.
Cependant, ces tables concernent des individus moyens et ne peuvent donc rendre compte des variations morphométriques des individus en fonction de leur environnement ou encore au cours de leur croissance.

L'idée de ce papier est de reconstruire la frontière de chacun des segments de l'individu modélisé
en utilisant les coordonnés cylindriques.
Fondamentalement, ce choix est guidé par le fait que chaque membre est proche d'un volume de révolution autour d'un axe 
et donc, qu'à altitude $z=z_0$ constante, la frontière de ce membre est proche d'un cercle, donc définie en coordonnées
polaires par 
la fonction $r$ : $\theta\mapsto r(\theta)$, où $r$ ne varie que <<~peu~>>.

Une autre idée consisterait à utiliser les fonctions $B$-splines comme dans \cite{Hutchinson2007}
(on pourra aussi consulter les ouvrages  \cite{martin2013,MR1900298}), où la frontière recherchée est définie 
par une surface $B$-spline. Cette méthode n'a pas été envisagée essentiellement pour deux raisons : 
\begin{itemize}
\item
Comme écrit dans \cite{Hutchinson2007} : 
<<Once the initial bounding volume is constructed, a set
of spatial points is generated uniformly over the volume \cite{NgThowHing1997}.
For each of these
spatial points, the closest data point is identified and the
$B$-spline solid is deformed to make the data point and
corresponding spatial point coincident. In general, this
method produces good initial shapes when the number of
control points are chosen to be fewer than the number of
data points. Subsequent manual adjustments of the control
points can be made for shape refinement.>>
L'absence de gestion automatique des points de contrôle qui définissent les splines impose
une intervention humaine et fastidieuse.
\item
Une spline est  définie par les points de contrôle et non par les points de passages, qui sont les seuls
expérimentalement connus. Un système linéaire permettrait de passer de l'un à l'autre et donc de définir la spline
grâce à la connaissance des points de passage. L'inversion de ce système linéaire risque cependant d'être 
source de longs calculs  et ce d'autant plus que les points expérimentaux de passage risquent d'être en nombre important.
\end{itemize}

\begin{figure}[h] 
\begin{center} 
\includegraphics[width=13 cm]{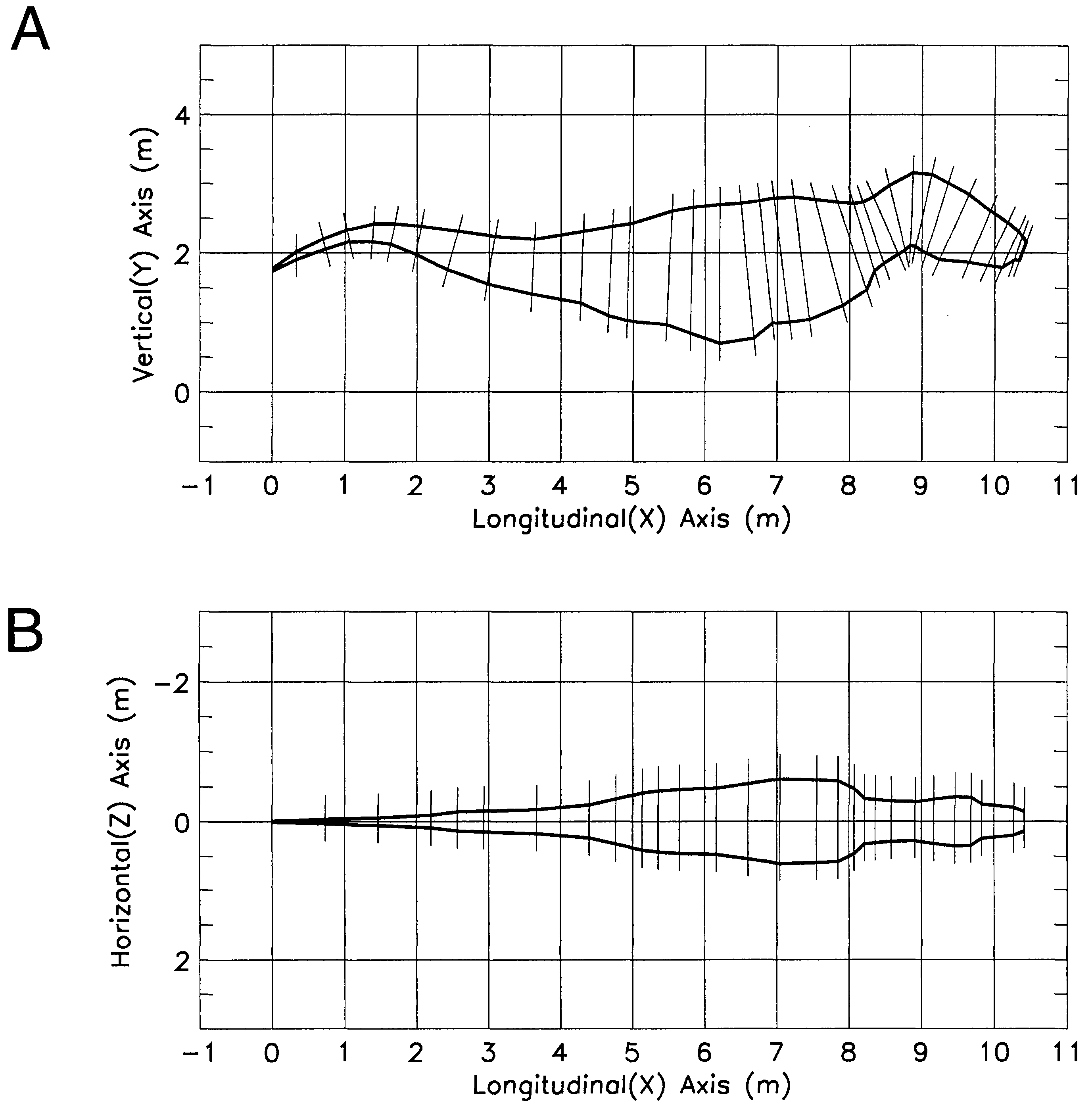}
\end{center} 
\caption{\label{Henderson_fig1}Recomposition 3D à partir de deux profils 2D.
Figure 1 de \cite{henderson1999}.} 
\end{figure}

Dans \cite{henderson1999}, Henderson utilise une technique dont la nôtre s'inspire, mais qui n'est pas satisfaisante en l'état. Elle a été adaptée et améliorée.
\`A partir de deux vues, l'une latérale et l'autre frontale (voir figure \ref{Henderson_fig1}, extraite de \cite{henderson1999}),
il détermine 4 points et leurs coordonnées, par mesure de l'intersection de segments avec le profil étudié. Il en déduit donc
ensuite une ellipse passant par ces quatre points et, grâce à l'ensemble des ces ellipses, il reconstitue le profil 3D, en en déduisant le calcul 
du centre de masse et du volume, en supposant la masse volumique constante. Cette méthode semble être apparemment utile pour reconstruire notre profil et en déduire
les données morphométriques et servira de point de départ ; cependant, on ne la conservera pas telle quelle pour plusieurs raisons : 
\begin{itemize}
\item
Cette méthode ne fonctionne que si l'animal étudié présente un plan sagittal de symétrie, ce qui ne sera pas toujours le cas, notamment si des ondulations sont présentes,
comme par exemple la queue de la figure \ref{exemple_lezard_photo_queue} ; 
\item
Seuls quatre points sont considérés et cette méthode ne permet pas d'en rajouter, ce qui pourra être fait pour celle présentée, pour prendre en compte 
certains détails ; 
\item
Enfin, dans cette méthode, seuls la position du centre et les volumes sont déterminés ; il manque les inerties, ce que l'on propose de déterminer.
\end{itemize}
Nous reviendrons néanmoins sur cette méthode que l'on comparera à celle proposée.

Enfin, une dernière idée aurait pu être utilisée, à l'instar des photos aériennes prises par l'IGN, 
la  stéréoscopie (voir \url{http://fr.wikipedia.org/wiki/Stéréoscopie}). 
Cette technique nécessite de faire plusieurs paires de photos, avec de légers décalages, afin de couvrir l'ensemble de la surface du membre étudié
et utilise ensuite des techniques de calculs plus sophistiquées.
Elle n'est pas pertinente ici, puisque l'on essaye de proposer une méthodologie recquérant peu de prises de vues et de mesures et des moyens 
de calcul relativement sommaires.
De plus, cette méthode ne permet pas d'exploiter le fait que, à $z$ fixé, la frontière est proche d'un cercle.

Nous présenterons en section
\ref{acquisition} l'acquisition des données (relevé géométriques des points de la frontière),
puis, en section 
\ref{reconstruction}, comment cette frontière est reconstruite, 
et enfin, en section
\ref{donneesmorphometriques}, le calcul des 
principales données morphométriques (volume, position du centre de masse et matrices d'inerties).
Puisque la masse volumique est supposée constantes, ce sont en fait des données géométriques qui sont calculées. 
Enfin, quelques simulations seront présentées en section 
\ref{simulation}.

%%%%%%%%%%%%%%%%%%%%%%%%%%%%%%%%%%%%%%%%%%%%%%%%%%%%%%%%%%%%%
%%%%%%%%%%%%%%%%%%%%%%%%%%%%%%%%%%%%%%%%%%%%%%%%%%%%%%%%%%%%%
\section{Acquisition des données}
\label{acquisition}

\begin{figure}
\centering
%%% sous figure 1
\subfigure[\label{fig01a}Vue de dos]
%{\epsfig{file=Pogona_dos.eps, width=8cm}}
{\includegraphics[width=8 cm]{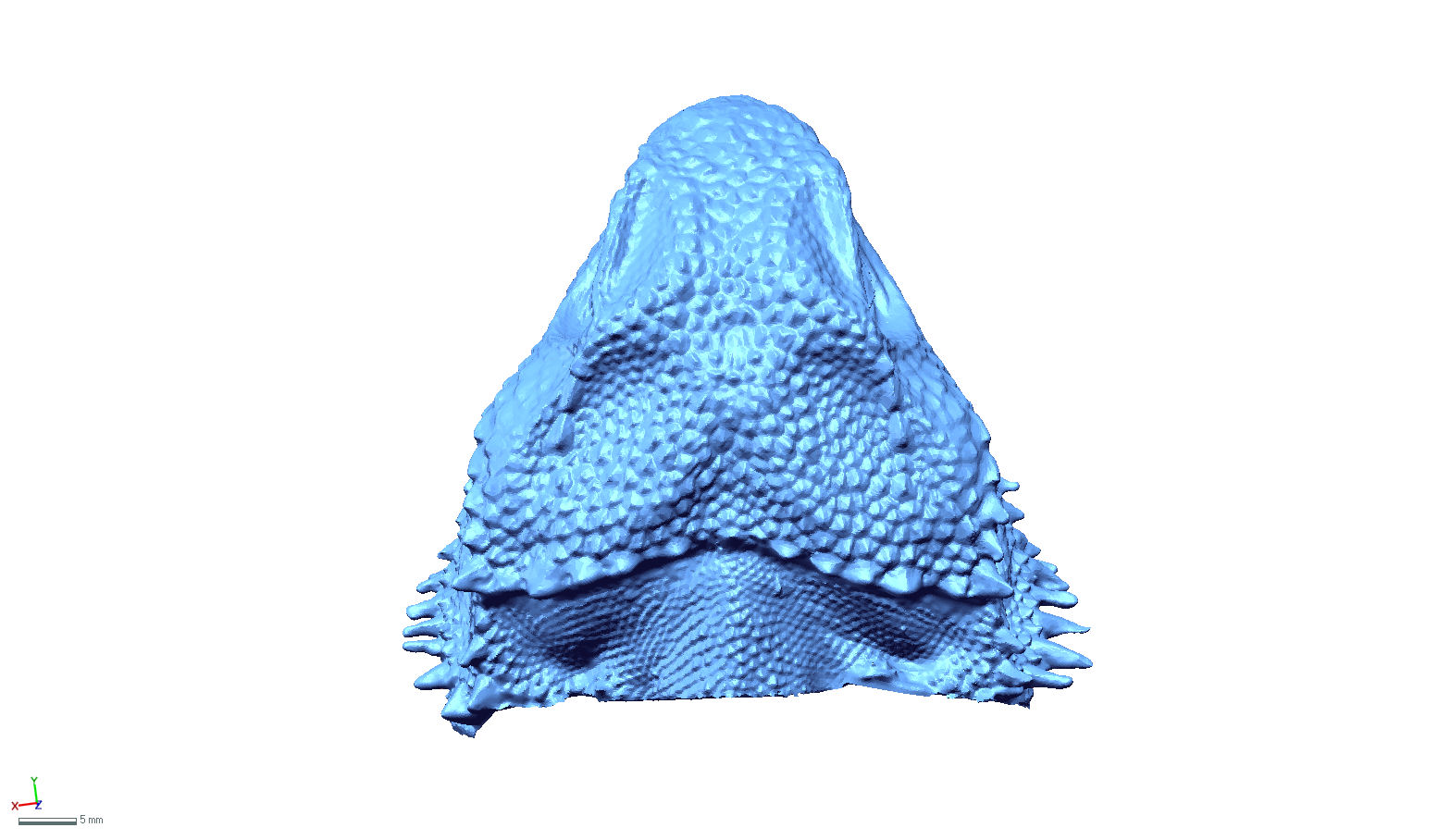}}
\qquad
%%% sous figure 2
\subfigure[\label{fig01b}Vue de profil]
%{\epsfig{file=Pogona_profil.eps, width=8cm}}
{\includegraphics[width=8 cm]{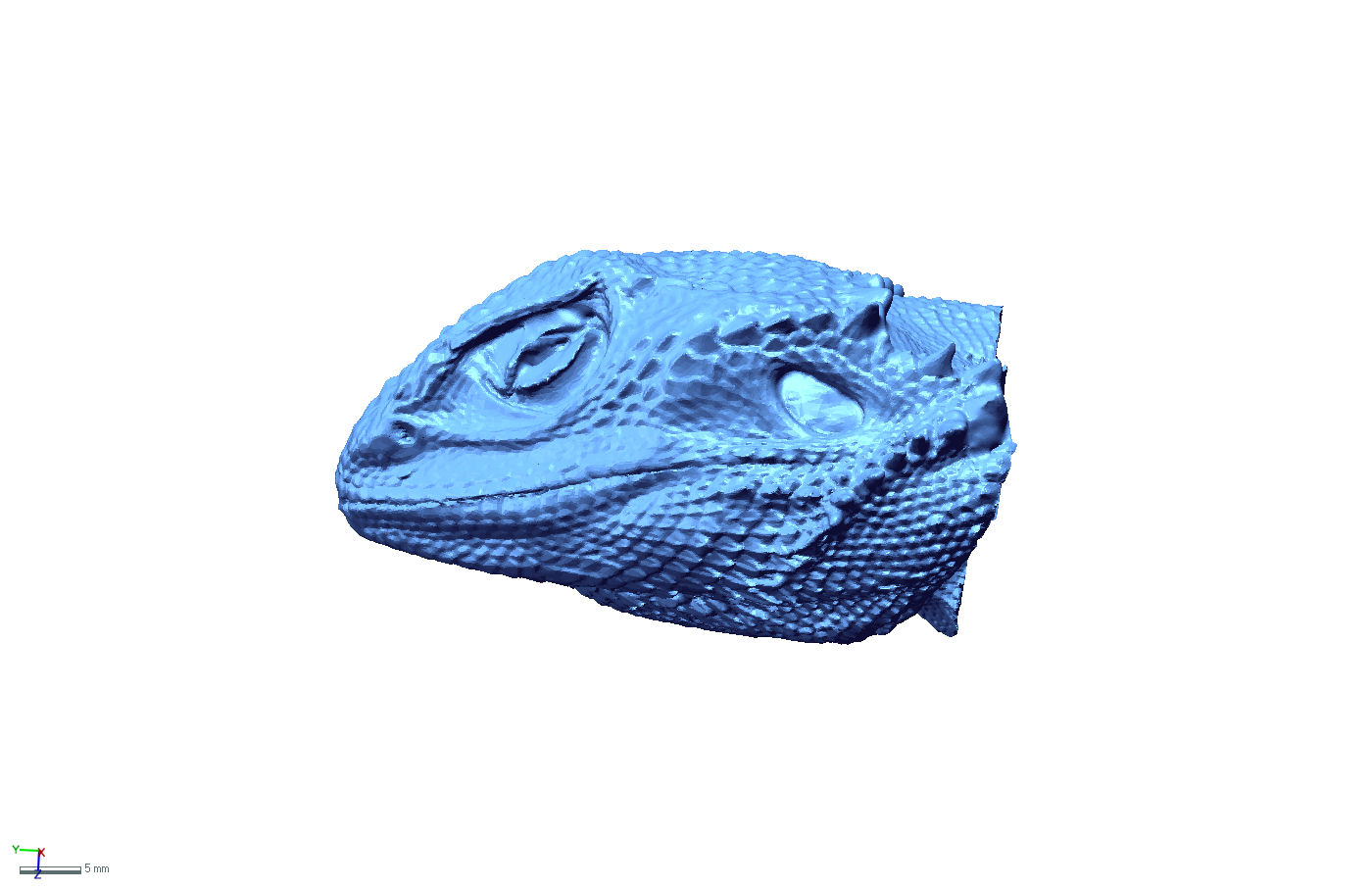}}
\caption{\label{fig01tot}Deux photos du Pogona.}
\end{figure}

On se donne deux photos de dos et de profil du membre considéré (ici la tête).
Voir figure \ref{fig01tot}.

\begin{figure}[h] 
\begin{center} 
\includegraphics[width=13 cm]{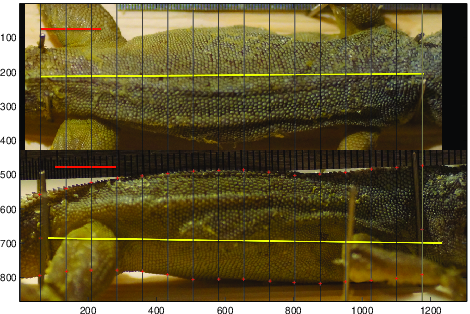}
\end{center} 
\caption{\label{fig01new}Deux photos du tronc et le repérage des points et des coupes.} 
\end{figure}

Chaque segment est digitalisé selon un nombre de coupes $n$  définies par l'utilisateur. Il en résulte $N=n+1$ plans de coupe. Le plan de coupe pour la vue de profil correspond au plan $(y,z)$. 
Le plan de coupe pour la vue dorsale correspond au plan $(x,z)$. L'axe $z$ correspond à l'axe longitudinal du segment 
(voir l'image \ref{fig01new}).
Les valeurs de $z$ sont donc celles des distances entre l'origine (fixée au premier plan de coupe à l'extrémité distale ou caudale du segment) et le plan de coupe considéré. Après digitalisation, toutes les coordonnées sont normalisées par la distance entre les plans de coupe des extrémités proximales et distales du segment. 

Dans les cas présentés par la suite, seuls quatre points symétriques ont été considéres comme dans \cite{henderson1999},
de telle sorte que l'on aurait pu utiliser les ellipses pour reconstituer notre volume. En fait, nous procéderons autrement, ce qui permettra un ajout possible de points.
Nous montrerons néanmoins que nos résultats sont très proches de ceux obtenus par  \cite{henderson1999} en section \ref{comaparaisonavechenderson}.

Pour chaque coupe d'altitude $(z_i)_{1\leq i \leq N}$,
on connaît donc les coordonnées $(x_j,y_j,z_i)_{(1\leq j \leq n_i)}$ de points de la  frontière.
On connaît donc un nombre $P=\sum_i n_i$ de points de la frontière du membre considéré,
notés aussi $(x_k,y_k,z_k)_{1\leq k \leq P}$.

Le tronc de l'image  \ref{fig01new} sera utilisé tout au long du papier ou du transparents comme
fil conducteur qui illustre la méthode.

\begin{remark}
\label{rem0m1}
Cette façon de procéder permet aussi d'enrichir les données en rajoutant des points caractéristiques, visibles sur les deux clichés.
\end{remark}

\begin{remark}
\label{rem00}
Chaque section sera mesurée soit par rapport à un centre correspondant à une droite fixée,
soit par rapport au  barycentre des points expérimentaux mesurés.
\end{remark}

\begin{remark}
\label{rem02}
Les différentes longueurs obtenues sont telles que l'altitude mimimale est nulle et l'altitude
maximale est égale  à 1. Il faudra donc multiplier toutes les longeurs obtenues
par un facteur d'échelle $\zeta$, égal par exemple (pour le tronc de la figure \ref{fig01new})  à 
%=15*0.004205 % ici 15=nombre de coupe demandée, 0.004205=largeur de coupe
$0.063075$
pour obtenir les longueurs en mètres.
Ainsi,  
les différentes  données morphométriques présentées en section \ref{donneesmorphometriques}
doivent être elles aussi multipliées 
respectivement par 
$\zeta^3$ pour obtenir le volume réel,
$\zeta$ pour les coordonnées du centre de masse,
et par $\zeta^5$ pour la matrice d'inertie.
\end{remark}

%%%%%%%%%%%%%%%%%%%%%%%%%%%%%%%%%%%%%%%%%%%%%%%%%%%%%%%%%%%%%
%%%%%%%%%%%%%%%%%%%%%%%%%%%%%%%%%%%%%%%%%%%%%%%%%%%%%%%%%%%%%
\section{Reconstruction de la frontière}
\label{reconstruction}

\begin{figure}[h] 
% \psfrag{x}{$x$}
% \psfrag{y}{$y$}
% \psfrag{z}{$z$}
% \psfrag{zm}{$z_{\min}$}
% \psfrag{zM}{$z_{\max}$}
% \psfrag{D}{$\mathcal{D}$}
% \psfrag{Dz}{$\partial \mathcal{D}(z)$}
% \psfrag{M}{$M$}
% \psfrag{th}{$\theta$}
% \psfrag{r}{$r(\theta,z)$}
\begin{center} 
\includegraphics[width=11 cm]{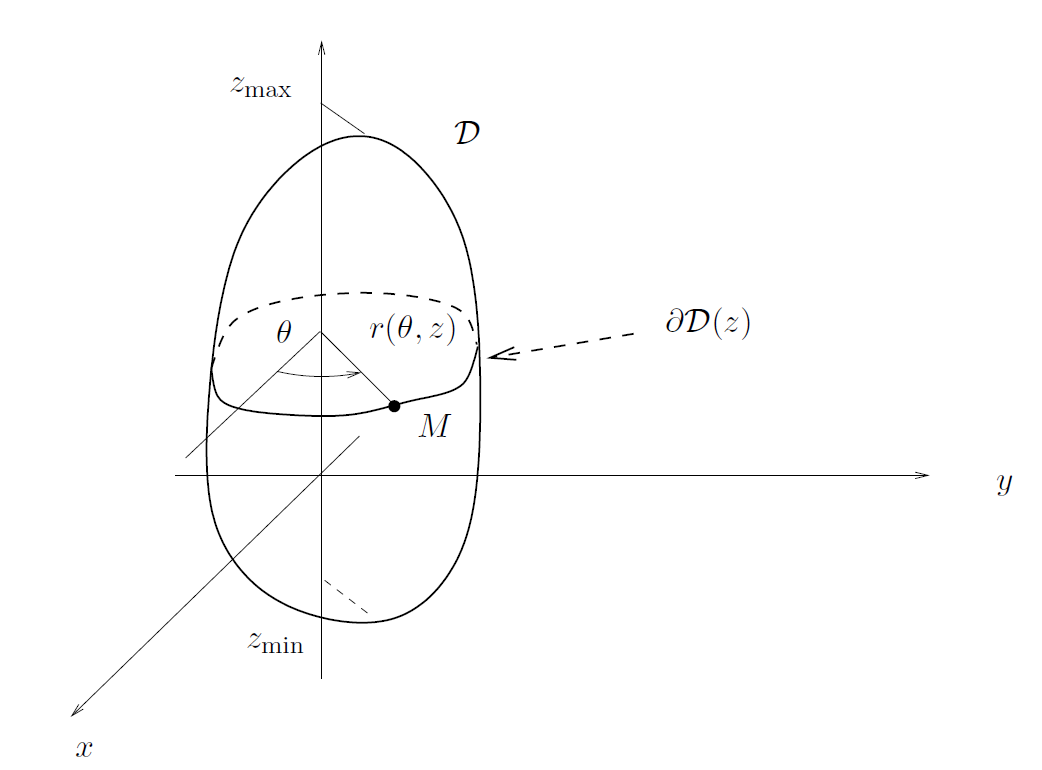}
\end{center} 
\caption{\label{fig01}Le volume $\mathcal{D}$.} 
\end{figure}

On se donne un volume $\mathcal{D}$ de l'espace, dont on suppose la  frontière $\partial \mathcal{D}$ connue sous la forme
de la surface paramétrée  donnée en coordonnées cylindrique par la 
fonction $r$ de $]-\pi,\pi]\times [z_{\min},z_{\max}]$ dans $\Er$ (voir figure \ref{fig01}) : 
\begin{equation}
\label{eq01}
(\theta,z) \mapsto r(\theta,z).
\end{equation}
On note $P_z$ le plan d'altitude $z$.
Ainsi, pour $z\in [z_{\min},z_{\max}]$ fixé,
la frontière $\partial \mathcal{D}(z)=\partial \mathcal{D}\cap P_z$
est décrite en coordonnées polaires par
\begin{equation}
\label{eq10}
\partial \mathcal{D}(z)=\Bigl\{M(x,y,z), \quad x=r(\theta,z)\cos\theta, \quad y=r(\theta,z)\sin\theta,\quad \theta\in ]-\pi,\pi] \Bigr\}.
\end{equation}

\begin{remark}\
\label{rem01}
\begin{figure}[h] 
% \psfrag{r}{$r$}
% \psfrag{th}{$\theta$}
% \psfrag{M}{$M(x,y,z)$}
% \psfrag{x0}{$x^0(z)$}
% \psfrag{x}{$x$}
% \psfrag{y}{$y$}
% \psfrag{y0}[][r]{$y^0(z)$}
\begin{center} 
\includegraphics[width=8 cm]{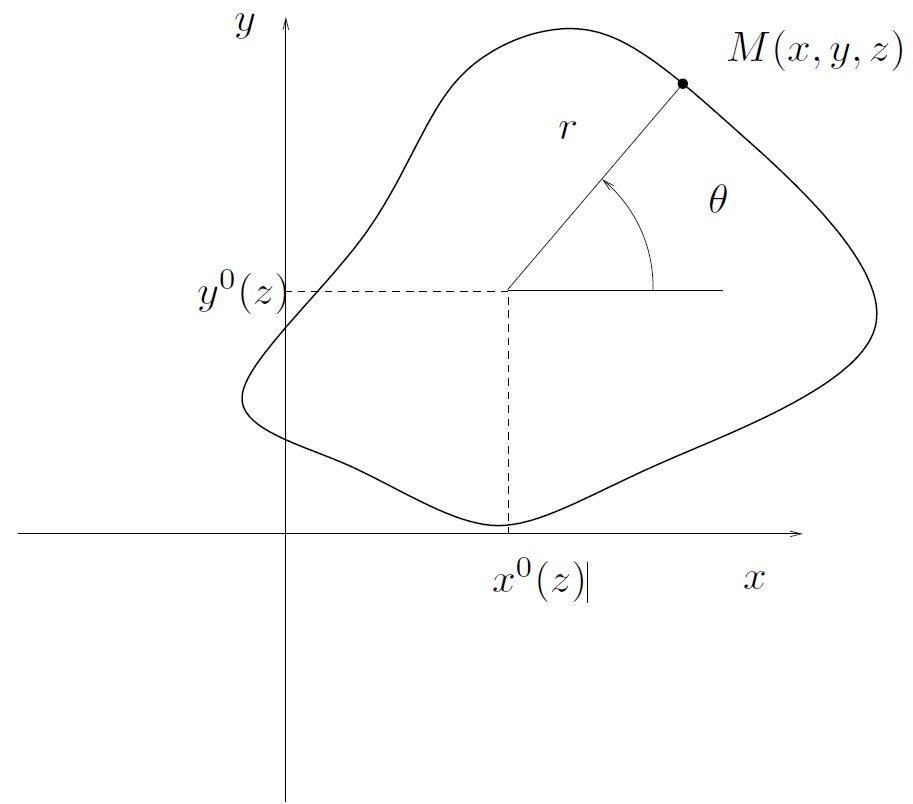}
\end{center} 
\caption{\label{fig02}La section $\mathcal{D}(z)$ et son centre $(x^0(z),y^0(z))$.} 
\end{figure}
Conformément à la remarque \ref{rem00}, on peut autoriser un centre variable de coordonnées $x^0(z)$ et $y^0(z)$ (dépendant de $z$) pour le repérage 
des coordonnées par section (voir figure \ref{fig02}) : on remplacera donc le paramétrage polaire \eqref{eq10}
par 
\begin{equation}
\label{eq11}
\partial \mathcal{D}(z)=\Bigl\{M(x,y,z), \quad x=r(\theta,z)\cos\theta+x^0(z), \quad y=r(\theta,z)\sin\theta+y^0(z),\quad \theta\in ]-\pi,\pi] \Bigr\}.
\end{equation}
Les fonctions $x^0$ et $y^0$ seront définies (constantes) par l'utilisateur ou mesurées expérimentalement.
\end{remark}

On cherche donc une fonction  $r$ et deux fonctions $x^0$ et $y^0$  vérifiant
\begin{subequations}
\label{eq20tot}
\begin{align}
\forall k \in \{1,\hdots,P\},\quad
\exists  \theta_k\in  ]-\pi,\pi],
\quad 
&x_k=r(\theta_k,z_k)\cos\theta_k+x^0(z_k),\\
&y_k=r(\theta_k,z_k)\sin\theta_k+y^0(z_k).
\end{align}
\end{subequations}
Ce problème est équivalent à 
\begin{subequations}
\label{eq30tot}
\begin{align}
\forall i \in \{1,\hdots,N\},
\quad 
\forall j \in \{1,\hdots,n_i\},
\quad 
\exists  \theta_{i,j}\in  ]-\pi,\pi],
\quad 
&x_j=r(\theta_{i,j},z_i)\cos\theta_{i,j}+x^0(z_i),\\
&y_j=r(\theta_{i,j},z_i)\sin\theta_{i,j}+y^0(z_i).
\end{align}
\end{subequations}

Nous allons proposer plusieurs méthodes et conserver celle qui semblera la mieux adaptée au problème donné.

%%%%%%%%%%%%%%%%%%%%%%%%%%%%%%%%%%%%%%%%%%%%%%%%%%%%%%%%%%%%%
%%%%%%%%%%%%%%%%%%%%%%%%%%%%%%%%%%%%%%%%%%%%%%%%%%%%%%%%%%%%%
% POUR INFO, extrait de 
% \programmes\matlab\recherche\CRIS\reconstruction_anatomie_interp\Morpho\principalA.m
% 
% reconstitution de la frontière et calculs des centres de masses, des
% volumes et des inerties.
%
% FINIR TOUS LES AAA et BBB
%
% *********************************************************
% [vertsb,vertsfinA,vertsfinB,vertsfinC,vertsfinD,VA,VB,VC,VD,CGA,CGB,CGC,CGB,IA,IB,IC,ID]=...
%    principalA(verts,centresect,trace2D,trace3D,tracesect,Nz,Nthe,nompdf,method)
%
% reconstitution de la frontière et calculs des centres de masses, des
% volumes et des inerties selon 4 méthodes :
%   - méthode A : reconstitution par coupe en interpolant dans
%     chaque coupe r(theta,z) en fonction de theta avec perpchip (voir
%     cette fonction)
%   - méthode B : reconstitution par coupe en interpolant dans
%     chaque coupe r(theta,z) en fonction de theta avec perspline (voir
%     cette fonction)
%   - méthode C : reconstitution par coupe en interpolant dans
%     chaque coupe r(theta,z) en fonction de theta avec des points ou r'(theta)
%     est imposée nulle (en supposant que ce sont des points extrémaux)
%   - méthode D : reconstitution en faisant un interpolation 2D de r(theta,z)
%%%%%%%%%%%%%%%%%%%%%%%%%%%%%%%%%%%%%%%%%%%%%%%%%%%%%%%%%%%%%
%%%%%%%%%%%%%%%%%%%%%%%%%%%%%%%%%%%%%%%%%%%%%%%%%%%%%%%%%%%%%

%%%%%%%%%%%%%%%%%%%%%%%%%%%%%%%%%%%%%%%%%%%%%%%%%%%%%%%%%%%%%
\subsection{Reconstruction de la frontière par coupe}
\label{reconstruction_coupe}

Le problème  \eqref{eq30tot} peut se découpler ainsi : 
pour chaque $i \in \{1,\hdots,N\}$, on cherche $x^0_i$, $y^0_i$ et une fonction $r_i$
de $]-\pi,\pi]$ dans $\Er$ 
vérifiant 
\begin{subequations}
\label{eq40tot}
\begin{align}
\forall j \in \{1,\hdots,n_i\},
\quad 
\exists  \theta_j\in  ]-\pi,\pi],
\quad 
&x_j=r_i(\theta_j)\cos\theta_{j}+x^0_i,\\
&y_j=r_i(\theta_j)\sin\theta_{j}+y^0_i.
\end{align}
\end{subequations}
Les fonctions $r$, $x^0$ et $y^0$ seront  alors connues 
uniquement aux points d'altitude $z_i$ avec
\begin{subequations}
\label{eq50}
\begin{align}
\label{eq50a}
\forall \theta \in ]-\pi,\pi],\quad
&r(\theta,z_i)=r_i(\theta),\\
\label{eq50b}
&x^0(z_i)=x^0_i,\\
\label{eq50c}
&y^0(z_i)=y^0_i.
\end{align}
\end{subequations}
On détermine alors pour chaque $i \in \{1,\hdots,N\}$ et pour chaque $j \in \{1,\hdots,n_i\}$,
$r_{i,j}$ et $\theta_{j}$ tels que 
\begin{align*}
&x_j=r_{i,j}  \cos\theta_{j}+x^0_i,\\
&y_j=r_{i,j}  \sin\theta_{j}+y^0_i.
\end{align*}
Bref, on cherche donc à trouver une fonction $\rho$ de   $]-\pi,\pi]$ dans $\Er$ telle que,
pour $Q\in \En^*$  donné, 
${(\theta_l)}_{1\leq l \leq Q}\in ]-\pi,\pi]^Q$ et ${(R_l)}_{1\leq l \leq Q}\in \Erp ^Q$ donnés, on ait
\begin{equation}
\label{eq60}
\forall l\in \{1,\hdots,Q\},\quad
\rho(\theta_l)=R_l.
\end{equation}
Sans perte de généralité, on supposera les $\theta_l$ strictement croissants.
La fonction $\rho$ est naturellement $2\pi$-périodique : on peut donc rajouter l'angle $\theta_0=\theta_Q-2\pi$ 
et la valeur $R_0=R_Q$ et remplacer \eqref{eq60} par 
\begin{subequations}
\label{eq70}
\begin{align}
\label{eq70a}
& R_0=R_Q,\\
\label{eq70b}
&\forall l\in \{0,\hdots,Q\},\quad
\rho(\theta_l)=R_l.
\end{align}
\end{subequations}
Pour éviter la présence de points anguleux, on cherchera une fonction $\rho$ dérivable.
On est donc exactement dans le cas de l'interpolation périodique.
Voir l'annexe \ref{46580percubint_Jbastien_2014} et \cite{46580percubint_Jbastien_2014}.
On aura donc une fonction polynomiale cubique par morceaux. 

Deux solutions d'offrent à nous :
\begin{itemize}
\item
Soit, on cherche une fonction $\rho$ de classe $C^1$ uniquement,
mais qui respecte la monotonie des données,
ce que permet la fonction \path|perpchip|.
Cette méthode sera notée la méthode A.
La fonction $\rho(\theta)$ ne variant que <<~peu~>>, cette méthode sera à privilégier puisqu'elle 
limite les variations d'ampltitude de la fonction $\rho$.
\item
Soit, on cherche une fonction $\rho$ de classe $C^2$, ce que permet la fonction \path|perspline|.
Cette méthode sera notée la méthode B.
\end{itemize}

\begin{figure}[h] 
\begin{center} 
\includegraphics[width=6 cm]{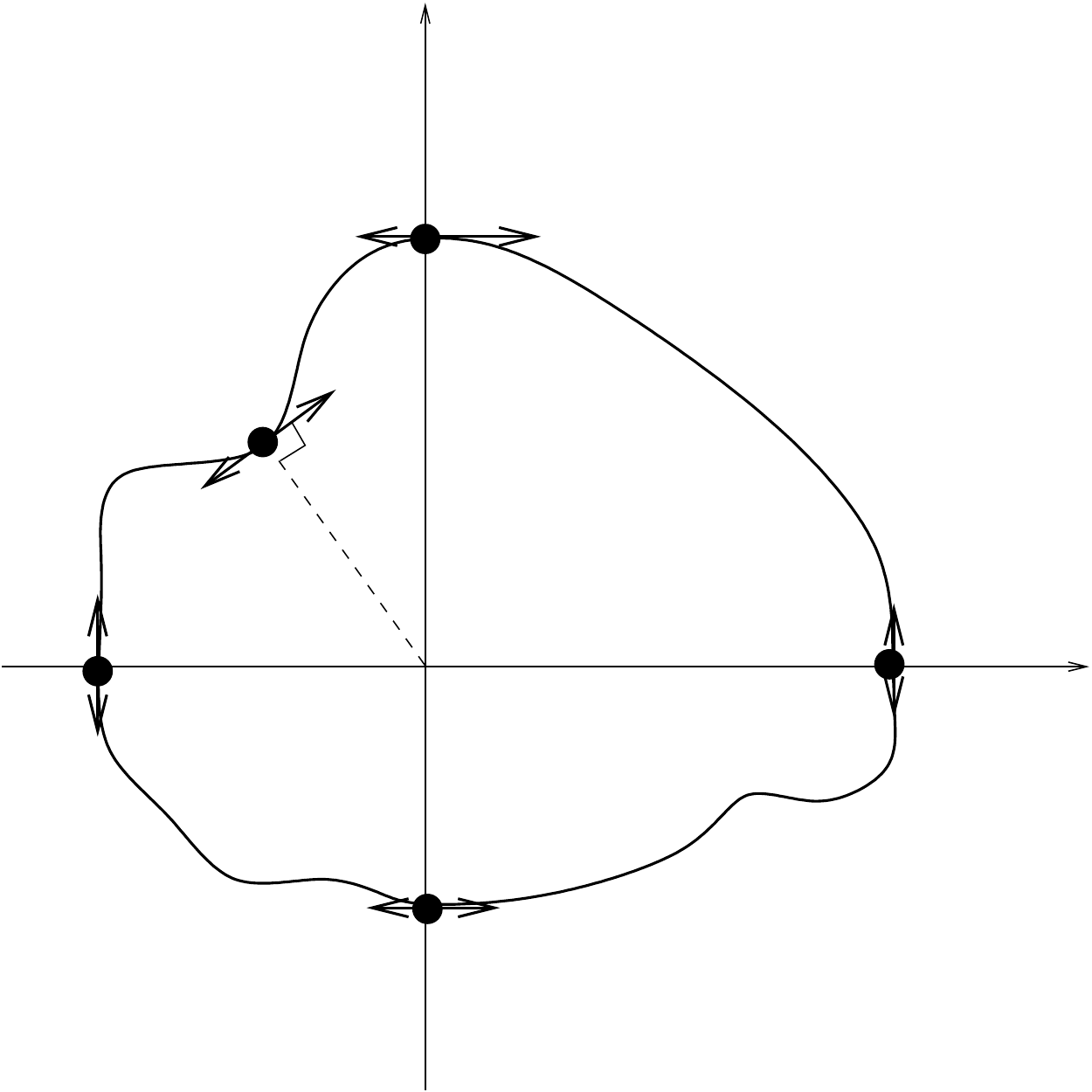}
\end{center} 
\caption{\label{fig10}Les points extrémaux considérés.} 
\end{figure}

Une troisième possibilité est aussi envisagée : on considère que les points mesurés correspondent à des extrémaux
locaux de $\rho$, ce qui est vérifié quand les points extrémaux sont considérés pour chaque profil :
voir figure \ref{fig10}.  On a donc la condition supplémentaire
\begin{equation}
\label{eq80}
\forall l\in \{0,\hdots,Q\},\quad
\rho'(\theta_l)=0.
\end{equation}
On cherchera donc l'unique fonction $\rho$ polynomiale cubique par morceaux sur chaque intervalle $[\theta_l,\theta_{l+1}]$,
de classe $C^1$ et vérifiant \eqref{eq70} et \eqref{eq80}.
Cette méthode sera notée la méthode C.

\begin{remark}
\label{methodcercle}
Une autre idée serait d'utiliser le fait qu'à $z$ fixé, la définition en polaire de
la frontière $\mathcal{D}(z)$ étant définie par  
la fonction $r$ : $\theta\mapsto r(\theta)$, où $r$ ne varie que <<~peu~>>.
On pourrait alors approcher cette 
frontière par un cercle, grâce à une méthode de moindre carré,
comme on a fait dans \cite{sigmo_TCJBCVPL_X1_12} :
\selectlanguage{english}
<<~The characteristics of the circle can be determined using a least squares method.
The center $\Omega$ and radius $R$ of the circle minimizing the distance sum of squares between experimental data  $(x_k,y_k)$ and theoretical data  $(X_k,Y_k)$ were researched. Firstly, a direct method was used to minimize the sum of 
\begin{equation*}
S={\left((x_k-X_k)^2+(y_k-Y_k)^2-R^2\right)}^2.
\end{equation*}
Secondly, the sum 
\begin{equation*}
S'=\sqrt{(x_k-X_k)^2+(y_k-Y_k)^2}
\end{equation*}
was minimized by using an iterative method. For this method, the results of the direct optimization were used as initial values for $\Omega$ and $R$. This final optimization was performed with the library \emph{M\MakeLowercase{ATLAB LEAST SQUARES GEOMETRIC ELEMENT SOFTWARE}},
available at 
{\tiny{\url{http://www.eurometros.org/gen_report.php?category=distributions&pkey=14}}}.
%\path|http://www.eurometros.org/gen_report.php?category=distributions&pkey=14|.
%\path|http://www.eurometros.org/gen_report.php?category=|\linebreak\path|distributions&pkey=14|.
For more details, see 
\cite{razet97,razet98,Ahnetali01,Endoetali07,MR1099915,MR1132365,MR1011392}.~>>

\selectlanguage{french}

Dans ce cas, $x^0_i$ et $y^0_i$ seront déterminés, ainsi que la fonction constante $\rho=\rho_i$.
Cette méthode n'a pas été implémentée car elle ne permet pas que la frontière passe exactement par chacun des points
expérimentaux.

\end{remark}

%%%%%%%%%%%%%%%%%%%%%%%%%%%%%%%%%%%%%%%%%%%%%%%%%%%%%%%%%%%%%
\subsection{Reconstruction de la frontière par interpolation à deux dimensions}
\label{reconstruction_2D}

On peut aussi considérer de nouveau le problème 
\eqref{eq20tot}. 
Dans le cas où $x^0$ et $y^0$ sont constants, on détermine,  pour chaque $k \in \{1,\hdots,P\}$, $R_k$ et $\theta_k$ tels que   
\begin{align*}
&x_k=R_k\cos\theta_{k}+x^0,\\
&y_k=R_k\sin\theta_{k}+y^0.
\end{align*}
On doit donc déterminer cette fois une fonction  $\rho$   de $]-\pi,\pi]\times [z_{\min},z_{\max}]$ dans $\Er$ telle que,
\begin{equation}
\label{eq100}
\forall k\in \{1,\hdots,P\},\quad
\rho(\theta_k,z_k)=R_k.
\end{equation}
On utilisera alors les deux fonctions 
\path|interp2| et \path|griddata| de matlab qui permettent de déterminer une telle fonction $\rho$,
sur une grille aussi fine que l'on veut et donc de calculer les valeurs de la frontière 
de $\mathcal{D}$.
Cette méthode sera notée la méthode D.

Cette méthode permet de déterminer des points de la frontière à toute altitude $z$, contrairement
aux méthodes A, B ou C, qui permet de  déterminer des points de la frontière uniquement aux altitudes $z_i$.
Mais,  elle est plus longue à calculer.  De plus, la périodicité en $\theta$ n'est plus assurée, ce qui fait apparaître des 
points anguleux sur la frontière reconstituée.

%%%%%%%%%%%%%%%%%%%%%%%%%%%%%%%%%%%%%%%%%%%%%%%%%%%%%%%%%%%%%
%%%%%%%%%%%%%%%%%%%%%%%%%%%%%%%%%%%%%%%%%%%%%%%%%%%%%%%%%%%%%
\section{Calculs des données morphométriques}
\label{donneesmorphometriques}

Nous déterminons dans cette section les principales données morphométriques (volume, position du centre de masse et matrices d'inerties),
grâce à la détermination de $\mathcal{D}$ et en supposant la masse volumique
$\mu$ de chacun des segments connue et homogène.

%%%%%%%%%%%%%%%%%%%%%%%%%%%%%%%%%%%%%%%%%%%%%%%%%%%%%%%%%%%%%
\subsection{Généralités et calcul en coordonnées cylindriques }
\label{gencylin}

Soit  une fonction $f$ est définie (et intégrable) de $\mathcal{D}$ dans $\Er$.
On considère les intégrales 
\begin{subequations}
\label{eq200tot}
\begin{align}
\label{eq200a}
&\widehat I=\iiint _\mathcal{D} f(x,y,z) dm,\\
\label{eq200b}
&I=\iiint _\mathcal{D} f(x,y,z) dV,\\
\end{align}
\end{subequations}
où $dm$ est l'élément de masse élémentaire autour du point de coordonnées $(x,y,z)$ et $dV$ le volume 
élémentaire autour de ce point, liés par
\begin{equation}
\label{eq210}
dm=\mu dV.
\end{equation}
Si $\mu$  est connue et homogène, on a donc
\begin{equation}
\label{eq215}
\widehat I=\mu I.
\end{equation}
Déterminons maintenant l'intégrale $I$.
On considère le changement de variable correspondant à \eqref{eq11}, c'est-à-dire, l'application 
qui à $(x,y,z)$ associe $(r,\theta,z)$ tels que, pour $z\in [z_{\min},z_{\max}]$ fixé, 
\begin{equation}
\label{eq216}
\begin{cases}
&x=r\cos\theta+x^0(z),\\
&y=r\sin\theta+y^0(z).
\end{cases}
\end{equation}
Les nouvelles variables sont $(r,\theta,z)$
et le jacobien $J=\frac{D(x,y,z)}{D(r,\theta,z)}$ est donné par  
\begin{equation*}
J=
\det
\begin{pmatrix}
\frac{\partial x}{\partial r} &\frac{\partial x}{\partial \theta} & \frac{\partial x}{\partial z}\\
\frac{\partial y}{\partial r} &\frac{\partial y}{\partial \theta} & \frac{\partial y}{\partial z}\\
\frac{\partial z}{\partial r} &\frac{\partial z}{\partial \theta} & \frac{\partial z}{\partial z}\\
\end{pmatrix}.
\end{equation*}
On a alors
\begin{equation*}
J=
\det
\begin{pmatrix}
\cos\theta & -r\sin\theta &{x^0}'\\
\sin\theta & r\cos\theta &{y^0}'\\
 0&0&1
\end{pmatrix}
=
r\cos^2\theta+r\sin^2\theta=r,
\end{equation*}
et on retrouve donc le fait que, comme en coordonnées  cylindriques, 
\begin{equation*}
dV=dxdydz=rdrd\theta dz.
\end{equation*}
On intègre donc, finalement, par tranches : 
\begin{align*}
I&=
 \iiint _{\substack{z\in [z_{\min},z_{\max}]  \\(x,y)\in \mathcal{D}(z)}} f(x,y,z) dV,\\
&=\int_{z=z_{\min}} ^{z_{\max}} \iint _{\substack{\theta\in [-\pi,\pi]\\r\in [0,r(\theta,z)]}}  f(r\cos\theta+x^0(z),r\sin\theta+y^0(z),z) rdrd\theta dz,
\end{align*}
Bref,
\begin{subequations}
\label{eq221tot}
\begin{align}
\label{eq221a}
&I=\int_{z_{\min}} ^{z_{\max}} F(z)dz,
\intertext{où}
\label{eq221b}
&F(z)=\iint _{\substack{\theta\in [-\pi,\pi]\\r\in [0,r(\theta,z)]}}  f(r\cos\theta+x^0(z),r\sin\theta+y^0(z),z) rdrd\theta.
\end{align}
\end{subequations}
Enfin, on suppose que l'on se place dans le cas particulier où  $A$, $B$ et $C$ sont trois entiers naturels,
que $\widehat x$,  $\widehat y$ et $\widehat z$ sont trois réels et que
et que
\begin{equation}
\label{eq230}
f(x,y,z)={\left(x-\widehat x\right)}^A {\left(y-\widehat y\right)}^B{\left(z-\widehat z \right)}^C
\end{equation}
On note 
\begin{equation}
\label{eq231p5}
I(A,B,C)=\int_{z_{\min}} ^{z_{\max}} {\left(z-\widehat z \right)}^C F(A,B,z)dz,
\end{equation}
et 
\begin{equation}
\label{eq231}
F(A,B,z)=
\iint _{\substack{\theta\in [-\pi,\pi]\\r\in [0,r(\theta,z)]}}
{\left(r\cos\theta+x^0(z)-\widehat x\right)}^A {\left(r\sin\theta+y^0(z)-\widehat y\right)}^B rdrd\theta.
\end{equation}

%%%%%%%%%%%%%%%%%%%%%%%%
\subsubsection{Calcul approché de l'intégrale fournissant $F$}\
\label{calculapproch}

\`A $z$ fixé, posons
\begin{subequations}
\label{eq222tot}
\begin{align}
\label{eq222a}
a=x^0(z)-\widehat x,\\
\label{eq222b}
b=y^0(z)-\widehat y.
\end{align}
\end{subequations}
On considère le polynôme à deux variables $P$ défini par 
\begin{equation}
\label{eq223}
P(X,Y)={\left(X+a\right)}^A{\left(Y+b\right)}^B.
\end{equation}
En le développant, on constate qu'il existe $N\in \En^*$,
${(c_i)}_{0\leq i \leq N}\in \Er^{N+1}$ (qui dépendent de $z$),
${(\alpha_i)}_{0\leq i \leq N}\in \Er^{N+1}$  et 
${(\beta_i)}_{0\leq i \leq N}\in \Er^{N+1}$ (qui ne dépendent que de $A$ et de $B$), tels que
\begin{equation}
\label{eq224}
P(X,Y)=\sum_{i=0}^N  c_iX^{\alpha_i}Y^{\beta_i}.
\end{equation}
Voir l'annexe \ref{calcul_poly} où les calculs de 
$N\in \En^*$,
${(c_i)}_{0\leq i \leq N} $, 
${(\alpha_i)}_{0\leq i \leq N} $ et 
${(\beta_i)}_{0\leq i \leq N} $ 
en fonction des valeurs particulières de $A$ et de $B$ utiles pour la suite sont donnés.
Ainsi, pour $z$ fixé, on a donc  
\begin{equation*}
{\left(r\cos\theta+x^0(z)-\widehat x\right)}^A {\left(r\sin\theta+y^0(z)-\widehat y\right)}^B
=
\sum_{i=0}^N  c_i{(r\cos\theta)}^{\alpha_i}{(r\sin\theta)}^{\beta_i},
\end{equation*}
et donc
\begin{equation*}
{\left(r\cos\theta+x^0(z)-\widehat x\right)}^A {\left(r\sin\theta+y^0(z)-\widehat y\right)}^B
=
\sum_{i=0}^N  c_i r^{\alpha_i+\beta_i}   \cos^{\alpha_i}\theta \sin^{\beta_i}\theta .
\end{equation*}
Ainsi, il vient
\begin{align*}
F(A,B,z)&=
\iint _{\substack{\theta\in [-\pi,\pi]\\r\in [0,r(\theta,z)]}}
{\left(r\cos\theta+x^0(z)-\widehat x\right)}^A {\left(r\sin\theta+y^0(z)-\widehat y\right)}^Brdrd\theta,\\
&=
\sum_{i=0}^N
\iint _{\substack{\theta\in [-\pi,\pi]\\r\in [0,r(\theta,z)]}}
c_i r^{\alpha_i+\beta_i}   \cos^\alpha_i\theta \sin^\beta_i\theta rdrd\theta,\\
&=
\sum_{i=0}^N
c_i
\iint _{\substack{\theta\in [-\pi,\pi]\\r\in [0,r(\theta,z)]}}
 r^{\alpha_i+\beta_i}   \cos^{\alpha_i}\theta \sin^{\beta_i}\theta
rdrd\theta,\\
&=
\sum_{i=0}^N
c_i
\int_{\theta=-\pi}^{\pi}
 \cos^{\alpha_i}\theta \sin^{\beta_i}\theta
\left(\int_{r=0}^{r(\theta,z)}
 r^{\alpha_i+\beta_i+1} 
dr\right)d\theta,
\intertext{et donc par intégration en $r$}
&=
\sum_{i=0}^N
\frac{c_i}{\alpha_i+\beta_i+2}
\int_{\theta=-\pi}^{\pi}
 \cos^{\alpha_i}\theta \sin^{\beta_i}\theta
r^{\alpha_i+\beta_i+2} (\theta,z)d\theta.
\end{align*}
Finalement, on a 
\begin{subequations}
\label{eq225}
\begin{align}
\label{eq225a}
&I(A,B,C)= \sum_{i=0}^N \int_{z_{\min}} ^{z_{\max}} {\left(z-\widehat z \right)}^C F_i(A,B,z)dz,
\intertext{où}
\label{eq225b}
&F_i(A,B,z)=
\frac{c_i}{\alpha_i+\beta_i+2}
\int_{-\pi}^{\pi}
\cos^{\alpha_i}\theta \sin^{\beta_i}\theta
r^{\alpha_i+\beta_i+2} (\theta,z)d\theta.
\end{align}
\end{subequations}
D'après la reconstruction de la frontière faite en section \ref{reconstruction}, on 
peut donc considérer un entier $M\in \En^*$ et ${(z_k)}_{1\leq k \leq M}$ (qui ne sont pas nécessairement les points expérimentaux)
vérifiant
\begin{equation}
\label{eq241}
z_{\min}\leq z_1\leq z_2\leq \hdots \leq z_M.
\end{equation}
On considère 
les pas ${(h_k)}_{1\leq k\leq M}$
 définis par
\begin{subequations}
\label{eq250tot}
\begin{align}
&h_1=z_2-z_1,\\
&\forall k\in \{2,\hdots,M-1\},\quad 
h_k=\frac{z_{k+1}-z_{k-1}}{2},\\
&h_M=z_M-z_{M-1}
\end{align}
\end{subequations}
On approche alors l'intégrale $I(A,B,C)$ de la façon suivante 
\begin{equation*}
I(A,B,C)\approx
\sum_{k=1}^M  
h_k {\left(z-\widehat z \right)}^C F(A,B,z_k),
\end{equation*}
et donc 
\begin{subequations}
\label{eq251}
\begin{align}
\label{eq251a}
&I(A,B,C)\approx
\sum_{i=0}^N 
\frac{1}{\alpha_i+\beta_i+2}
\sum_{k=1}^M
h_kI_{i,k}(z_k){(z_k-\widehat z)}^Cc_i(z_k),
\intertext{où}
\label{eq251b}
&
I_{i,k}=
\int_{-\pi}^{\pi}
\cos^{\alpha_i}\theta \sin^{\beta_i}\theta
r^{\alpha_i+\beta_i+2} (\theta,z_k)d\theta,
\end{align}
\end{subequations}
et $c_i(z_k)$ est défini par 
le tableau  \ref{calcul_poly} (où $a$ et $b$ sont définis en fonction de $z_k$
grâce à \eqref{eq222tot}).
Cette intégrale sera approchée par exemple par la méthode des trapèzes (voir par exemple \cite{jbjnmdunod03}).

%%%%%%%%%%%%%%%%%%%%%%%%
\subsubsection{Calcul exact de l'intégrale fournissant $F$}\
\label{calculexact}

La formule \eqref{eq231p5}-\eqref{eq231} peut se réécrire, compte tenu de \eqref{eq222tot}, pour chaque 
valeur de $z$ : 
\begin{equation}
\label{eq231P5tot}
F(A,B,z)=
\int_{\theta=-\pi} ^{\pi}
\int_{r=0} ^{r(\theta,z)}   
{\left(r\cos\theta+a\right)}^A {\left(r\sin\theta+b\right)}^B rdrd\theta.
\end{equation}
La détermination de la fonction $r(\theta,z)$
(à $z$ fixé, pour les méthodes A à C) 
se fait par l'intermédiaire de fonctions
qui fournissent un polynôme de degré 3 par morceaux.
On peut donc supposer que, sur chacun des intervalles considérés du type $[\theta_0,\theta_1]$, on a
\begin{equation}
\label{eq301}
\forall \theta \in [\theta_0,\theta_1],\quad
r(\theta,z)=
(\theta-\theta_0)^3p_3+
(\theta-\theta_0)^2p_2+
(\theta-\theta_0)p_1+p_0.
\end{equation}
Il faut donc déterminer, pour chacun de ces intervalles,
\begin{equation*}
\mathcal{G}(A,B,z,\theta_0,\theta_1)=
\int_{\theta=\theta_0} ^{\theta_1}
\left(
\int_{r=0}^{%
(\theta-\theta_0)^3p_3+
(\theta-\theta_0)^2p_2+
(\theta-\theta_0)p_1+p_0
}
{\left(r\cos\theta+a\right)}^A {\left(r\sin\theta+b\right)}^B rdr
\right)
d\theta.
\end{equation*}
Dans l'intégrale en $\theta$, on  fait le changement de variable 
$\theta'=\theta-\theta_0$. Il vient donc
\begin{equation}
\label{eq309}
\mathcal{G}(A,B,z,\theta_0,\theta_1)=
\int_{\theta=0} ^{\theta_1-\theta_0}
\left(
\int_{r=0}^{%
\theta^3p_3+
\theta^2p_2+
\theta  p_1+p_0
}
{\left(r\cos(\theta+\theta_0)+a\right)}^A {\left(r\sin(\theta+\theta_0)+b\right)}^B rdr
\right)
d\theta.
\end{equation}
On est capable de déterminer explicitement cette fonction (en utilisant par exemple matlab symbolique).
On a donc 
\begin{equation}
\label{eq319}
F(A,B,z)=
\sum_{l=0} ^{P-1}
\mathcal{G}(A,B,z,\theta_l,\theta_{l+1})
.
\end{equation}
On approchera finalement l'intégrale $I(A,B,C)$
comme précédemment : 
\begin{equation}
\label{eq330}
I(A,B,C)\approx
\sum_{k=1}^M
h_k{\left(z_k-\widehat z \right)}^C
F(A,B,z_k).
\end{equation}

%%%%%%%%%%%%%%%%%%%%%%%%
\subsubsection{Calcul exact de l'intégrale fournissant $F$ dans le cadre des moindres carrés}\
\label{calculexactmethodcercle}

Si on utilise la méthode de la remarque \ref{methodcercle},
le calcul est plus simple. 
On utilise de nouveau les formules \eqref{eq251}. Pour le calcul de $I_{i,k}$, on utilise le 
fait que $r(.,z_k)$  est constant et égal à $\rho_k$.
On a donc 
\begin{equation*}
I_{i,k}=
\rho_k^{\alpha_i+\beta_i+2}
\int_{-\pi}^{\pi}
\cos^{\alpha_i}\theta \sin^{\beta_i}\theta
d\theta,
\end{equation*}
cette intégrale pouvant se calculer explicitement (en utilisant matlab symbolique par exemple).

%%%%%%%%%%%%%%%%%%%%%%%%%%%%%%%%%%%%%%%%%%%%%%%%%%%%%%%%%%%%%
\subsection{Applications}

On choisit tout d'abord 
\begin{equation}
\label{eq252}
\widehat x=\widehat y=\widehat z=0.
\end{equation}
Le volume $V$ de $\mathcal{V}$ est donné par
\begin{equation*}
I=\iiint _\mathcal{D}  dV
\end{equation*}
et donc
donc finalement 
\begin{equation}
\label{eq300}
V=I(0,0,0).
\end{equation}
Par exemple, l'abcisse $x_G$ du centre de masse  de $\mathcal{V}$, de masse $\mathcal{M}$, est donnée par 
\begin{equation*}
x_G=\frac{1}{\mathcal{M}}\iiint _\mathcal{D}xdm=\frac{1}{\mu V} \iiint _\mathcal{D}x\mu dV
\end{equation*}
et donc, de façon plus générale, 
pour la position du centre de masse, on a 
\begin{subequations}
\label{eq310}
\begin{align}
&x_G=\frac{1}{V}I(1,0,0),\\
&y_G=\frac{1}{V}I(0,1,0),\\
&z_G=\frac{1}{V}I(0,0,1).
\end{align}
\end{subequations}
On choisit maintenant  
\begin{equation}
\label{eq253}
\widehat x=x_G,\quad
\widehat y=y_G,\quad
\widehat z=z_G.
\end{equation}
La matrice d'inertie est donnée par
\begin{equation*}
%\label{eq320ante}
I=
\mu
\begin{pmatrix}
\iiint _\mathcal{D}  {(y-y_G)}^2+ {(z-z_G)}^2dV& -\iiint _\mathcal{D}  (x-x_G)(y-y_G)     dV  & -\iiint _\mathcal{D}  (x-x_G)(z-z_G) dV\\
-\iiint _\mathcal{D}   (x-x_G)(y-y_G)  dV& \iiint _\mathcal{D} {(x-x_G)}^2+ {(z-z_G)}^2  dV  & -\iiint _\mathcal{D} (y-y_G)(z-z_G)  dV\\
-\iiint _\mathcal{D}  (x-x_G)(z-z_G) dV& -\iiint _\mathcal{D} (y-y_G)(z-z_G)  dV  & \iiint _\mathcal{D} {(x-x_G)}^2+ {(y-y_G)}^2  dV
\end{pmatrix}
\end{equation*}
 et il vient finalement 
\begin{equation}
\label{eq320}
I=
\mu
\begin{pmatrix}
 I(0,2,0)+ I(0,0,2) & - I(1,1,0) &- I(1,0,1) \\
 - I(1,1,0)  &  I(2,0,0)+ I(0,0,2)& - I(0,1,1) \\
- I(1,0,1)  & - I(0,1,1)&  I(2,0,0)+ I(0,2,0)
\end{pmatrix}
\end{equation}
Souvent, $\mu$ est de l'ordre de 1000. On normalisera en choisissant $\mu=1$  dans les simulations présentées plus loin.

\begin{remark}
\label{rem10}
Les inerties obtenues devront donc être multipliée par la masse volumique réelle.
Voir aussi la remarque \ref{rem02}.
\end{remark}

%%%%%%%%%%%%%%%%%%%%%%%%%%%%%%%%%%%%%%%%%%%%%%%%%%%%%%%%%%%%%
%%%%%%%%%%%%%%%%%%%%%%%%%%%%%%%%%%%%%%%%%%%%%%%%%%%%%%%%%%%%%
\section{Simulations}
\label{simulation}

%%%%%%%%%%%%%%%%%%%%%%%%%%%%%%%%%%%%%%%%%%%%%%%%%%%%%%%%%%%%%
\subsection{Validation : comparaison avec un ellipsoïde}
\label{simulation_compare_ellipsoide}

%%%%%%%%%%%%%%%%%%%%%%%%%%%%%%%%%%%%%%%%%%%%%%%%%%%%%%%%%%%%%
%\input{validation_ellipsoide}
% fichier tex crée par MaTeXBuild02 le 23-Jun-2014 11:43:32
% corrigé manuellement le 06/07/2015 à 16:48.
% à compiler avec clear all;MaTeXBuild02('validation_ellipsoide',0)
% ATTENTION, long 

Comme dans \cite[p. 96]{henderson1999}, nous avons comparé nos résultats avec les résultats exacts d'un ellipsoïde, d'équation
\begin{equation}
\label{valelieq01}
\frac{x^2}{a^2}+\frac{y^2}{b^2}+\frac{z^2}{c^2}\leq 1,
\end{equation}
où les trois demi-axes sont donnés. 
En plus de la comparaison avec le volume analogue à celle de \cite{henderson1999}, nous donnons aussi la comparaison
avec le centre de masse et les inerties.
On peut montrer que le volume de cet ellipsoïde est égal à 
\begin{equation}
\label{valelieq10}
V=\frac{4}{3}\pi a b c,
\end{equation}
tandis que le centre de gravité est nul et que 
\begin{equation}
\label{valelieq20}
I=
\frac{4}{15}\pi a b c 
\begin{pmatrix}
b^2+c^2 & 0& 0\\
0 & a^2+c^2 & 0\\
0 & 0 & a^2+b^2
\end{pmatrix}
\end{equation}

Comme dans  \cite{henderson1999},
nous avons fait varier 
$N-1\in 
\{2,4,8,16,32\}$ le nombre total de coupes et 
$Q\in 
\{4,8,16,32\}$, le nombre total de valeurs d'angles choisis.
Nous avons aussi choisi
$a=1.0$,
$b=1.6$ et 
$c=2.4$.

\begin{figure}
\centering
%%% sous figure 
\subfigure[\label{quelquesvolumes_N3_Q4} : $N=3$ et $Q=4$]
%{\epsfig{file=quelquesvolumes_N3_Q4.eps, width=5cm}}
{\includegraphics[width=5 cm]{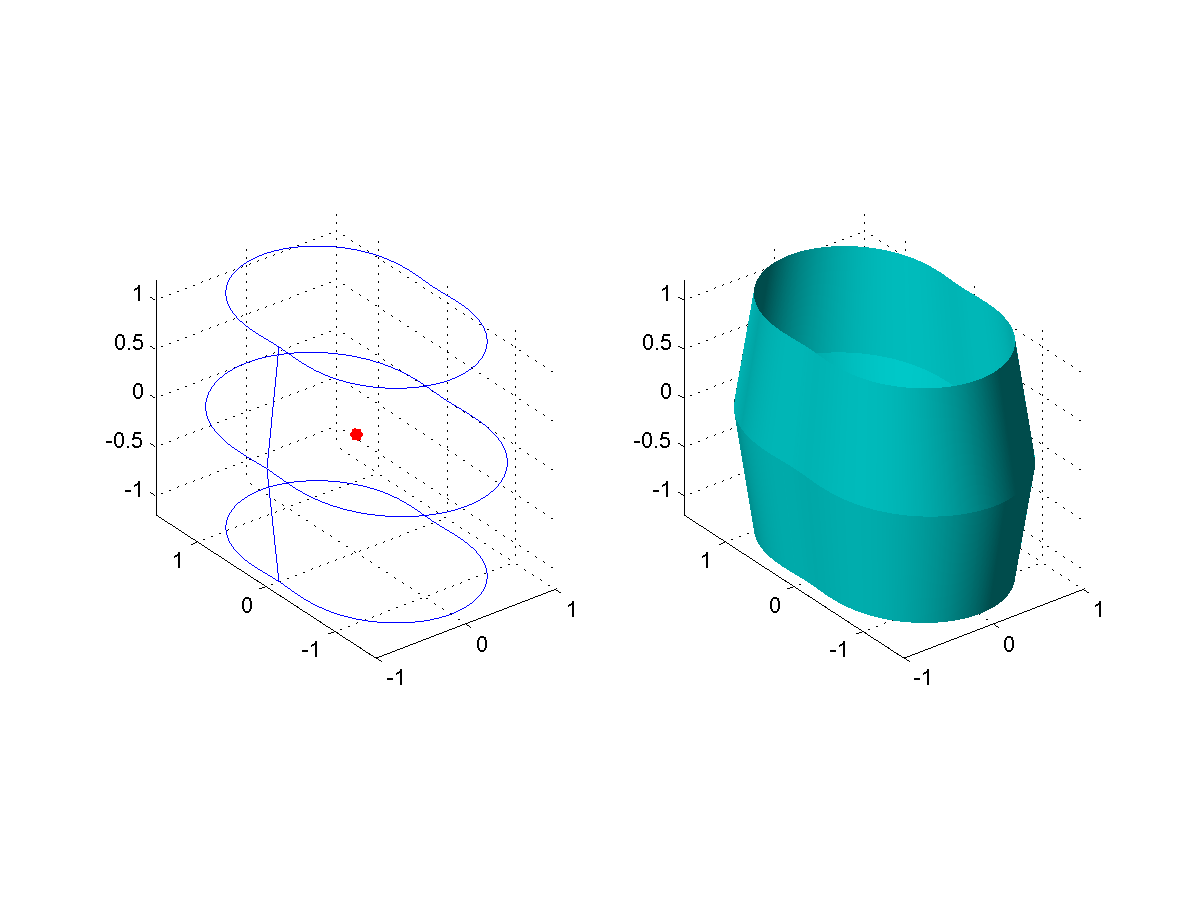}}
%%% sous figure 
\subfigure[\label{quelquesvolumes_N3_Q16} : $N=3$ et $Q=16$]
%{\epsfig{file=quelquesvolumes_N3_Q16.eps, width=5cm}}
{\includegraphics[width=5 cm]{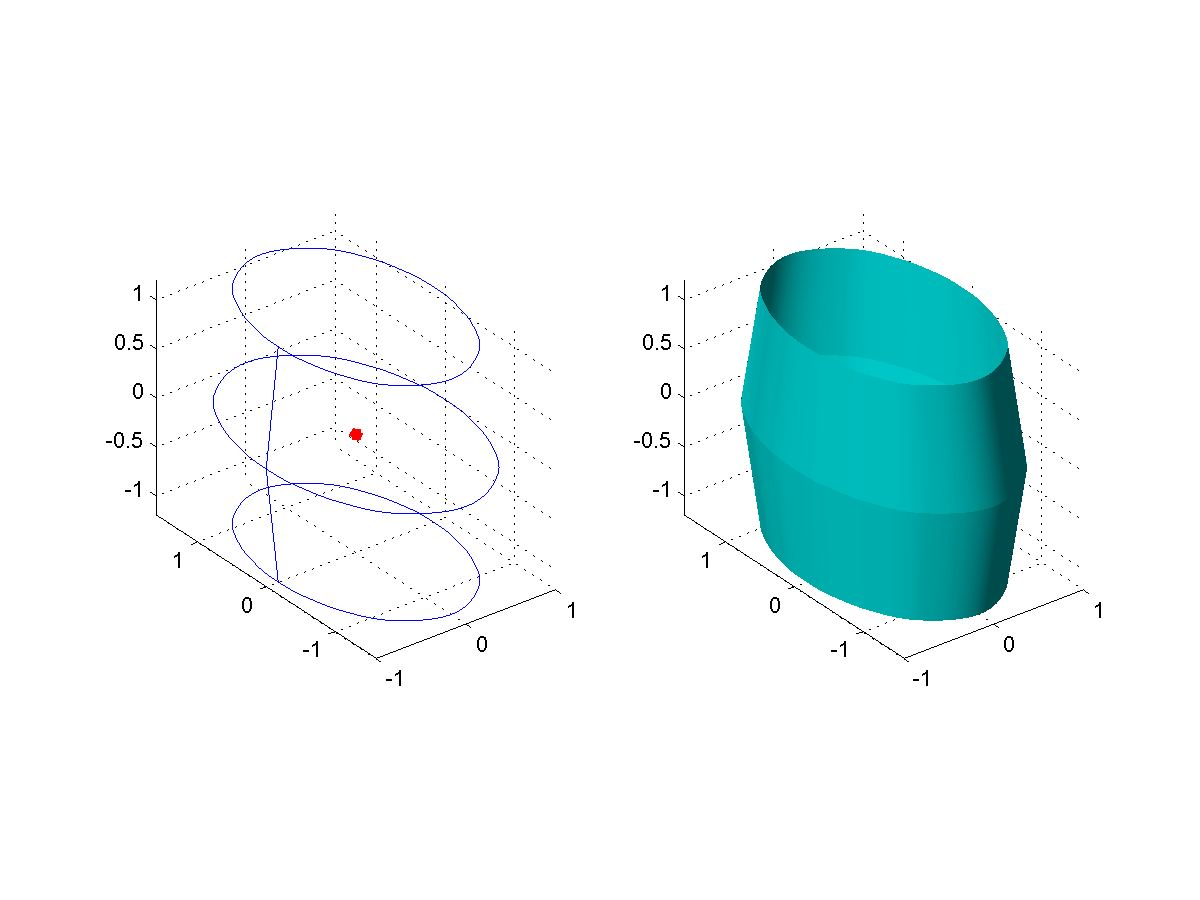}}
%%% sous figure 
\subfigure[\label{quelquesvolumes_N3_Q32} : $N=3$ et $Q=32$]
%{\epsfig{file=quelquesvolumes_N3_Q32.eps, width=5cm}}
{\includegraphics[width=5 cm]{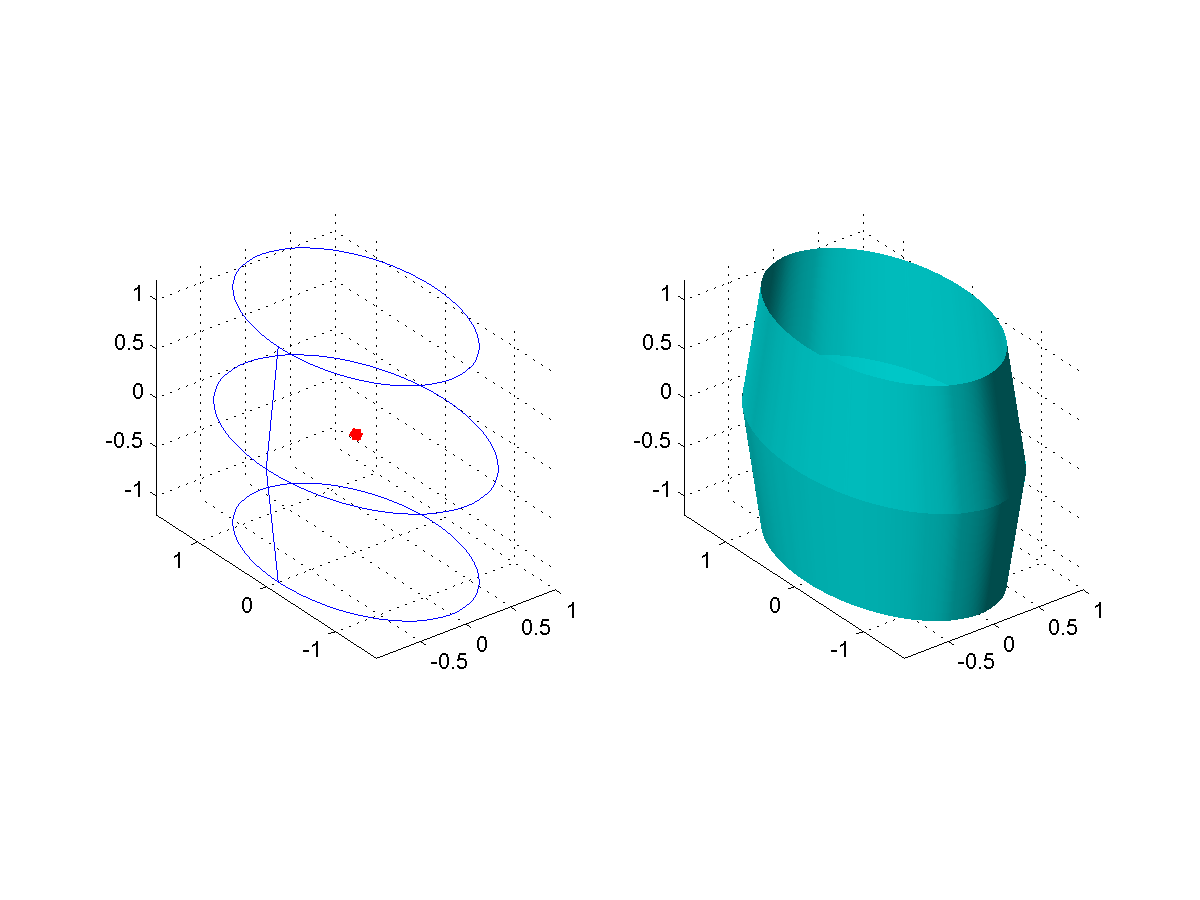}}
%%% sous figure 
\subfigure[\label{quelquesvolumes_N9_Q4} : $N=9$ et $Q=4$]
%{\epsfig{file=quelquesvolumes_N9_Q4.eps, width=5cm}}
{\includegraphics[width=5 cm]{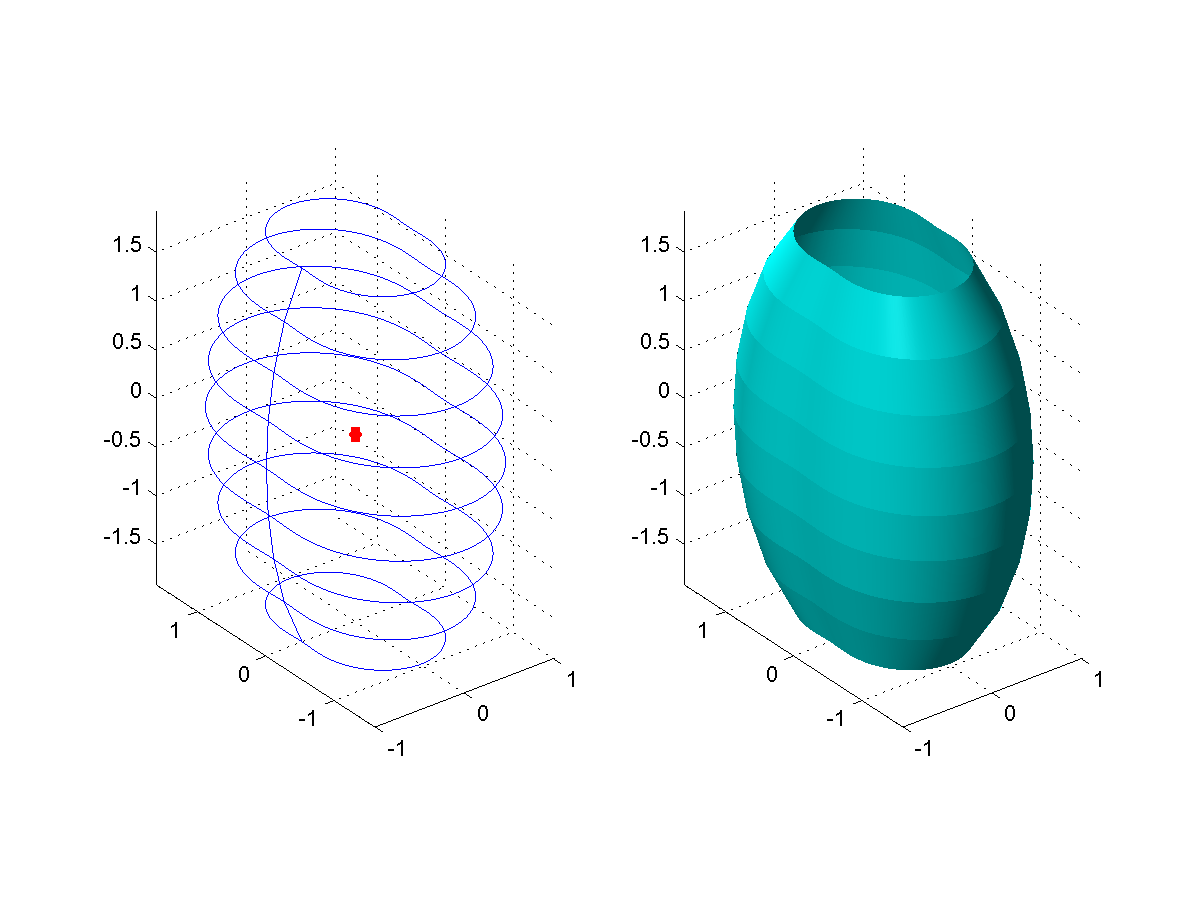}}
%%% sous figure 
\subfigure[\label{quelquesvolumes_N9_Q16} : $N=9$ et $Q=16$]
%{\epsfig{file=quelquesvolumes_N9_Q16.eps, width=5cm}}
{\includegraphics[width=5 cm]{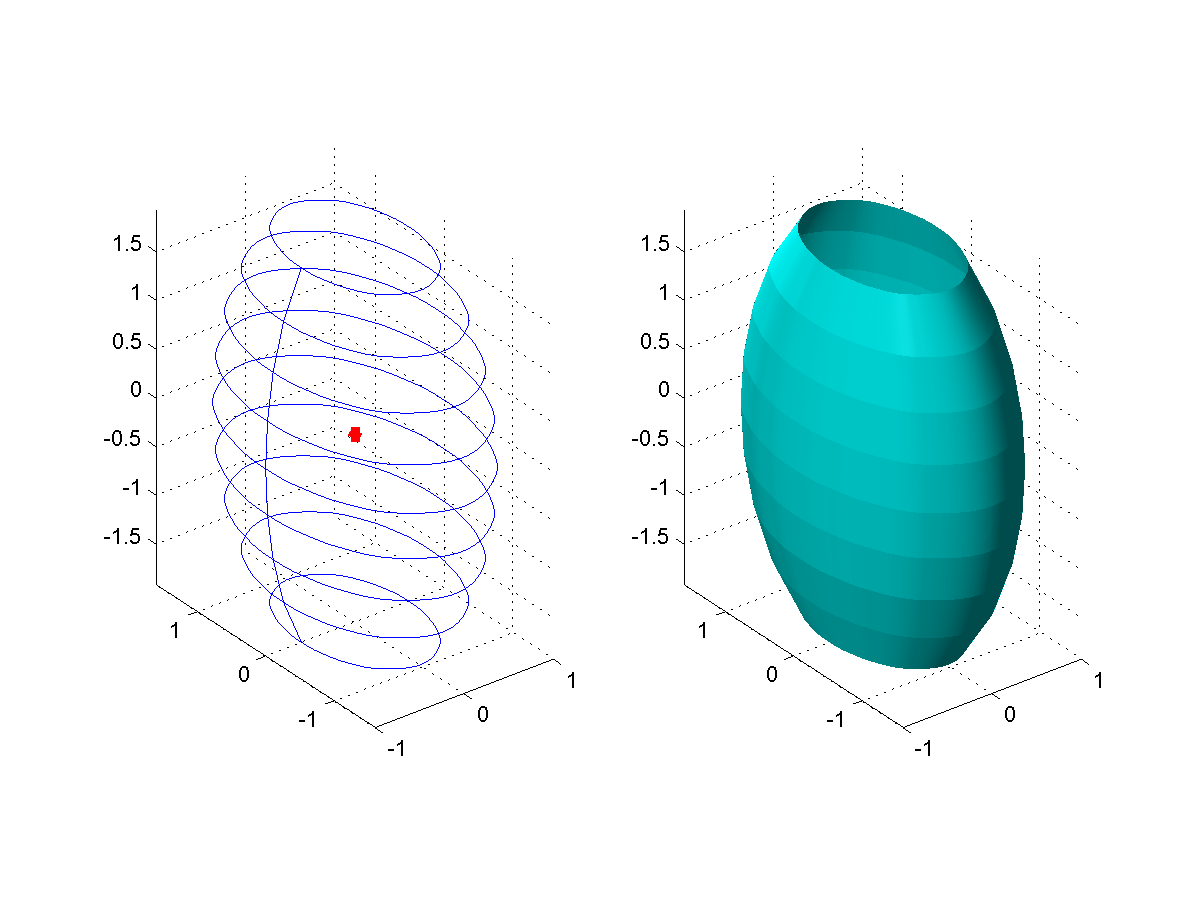}}
%%% sous figure 
\subfigure[\label{quelquesvolumes_N9_Q32} : $N=9$ et $Q=32$]
%{\epsfig{file=quelquesvolumes_N9_Q32.eps, width=5cm}}
{\includegraphics[width=5 cm]{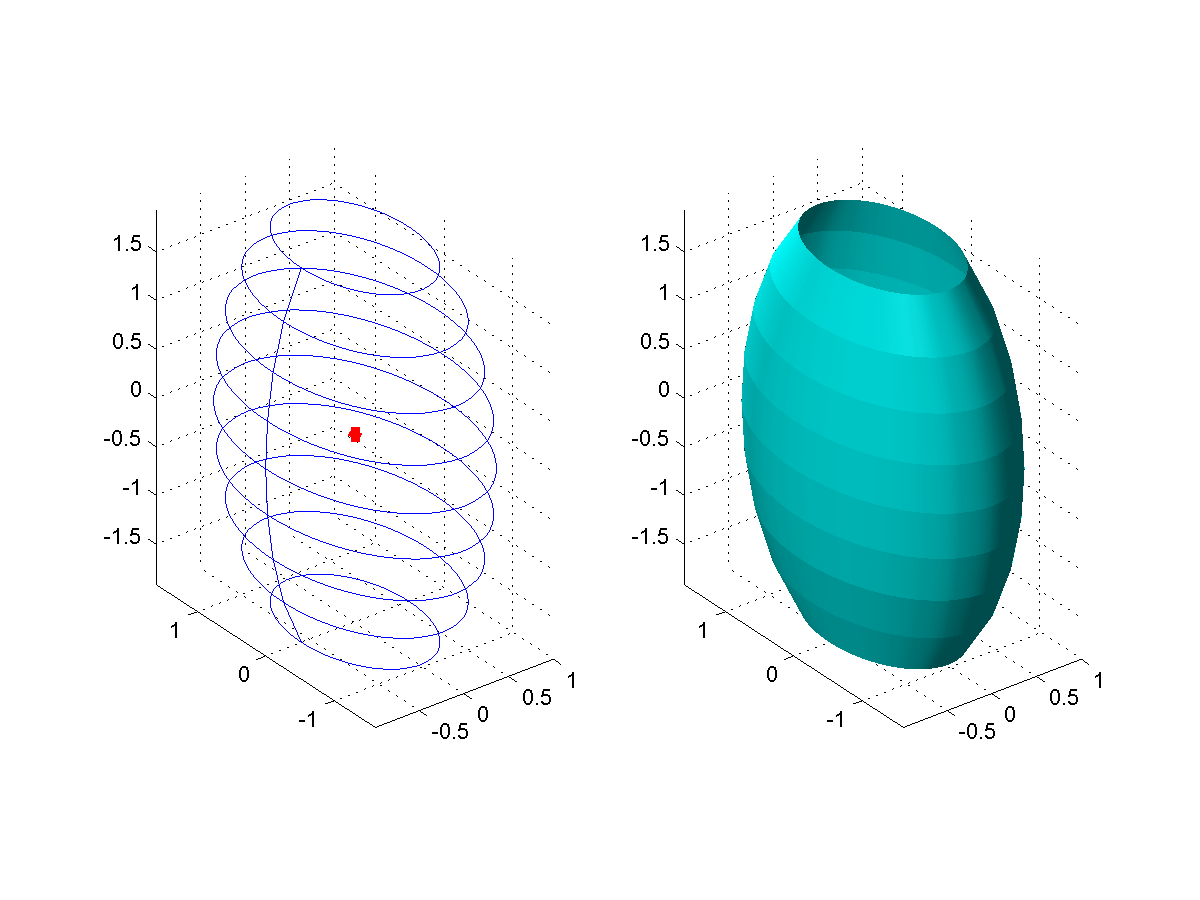}}
%%% sous figure 
\subfigure[\label{quelquesvolumes_N33_Q4} : $N=33$ et $Q=4$]
%{\epsfig{file=quelquesvolumes_N33_Q4.eps, width=5cm}}
{\includegraphics[width=5 cm]{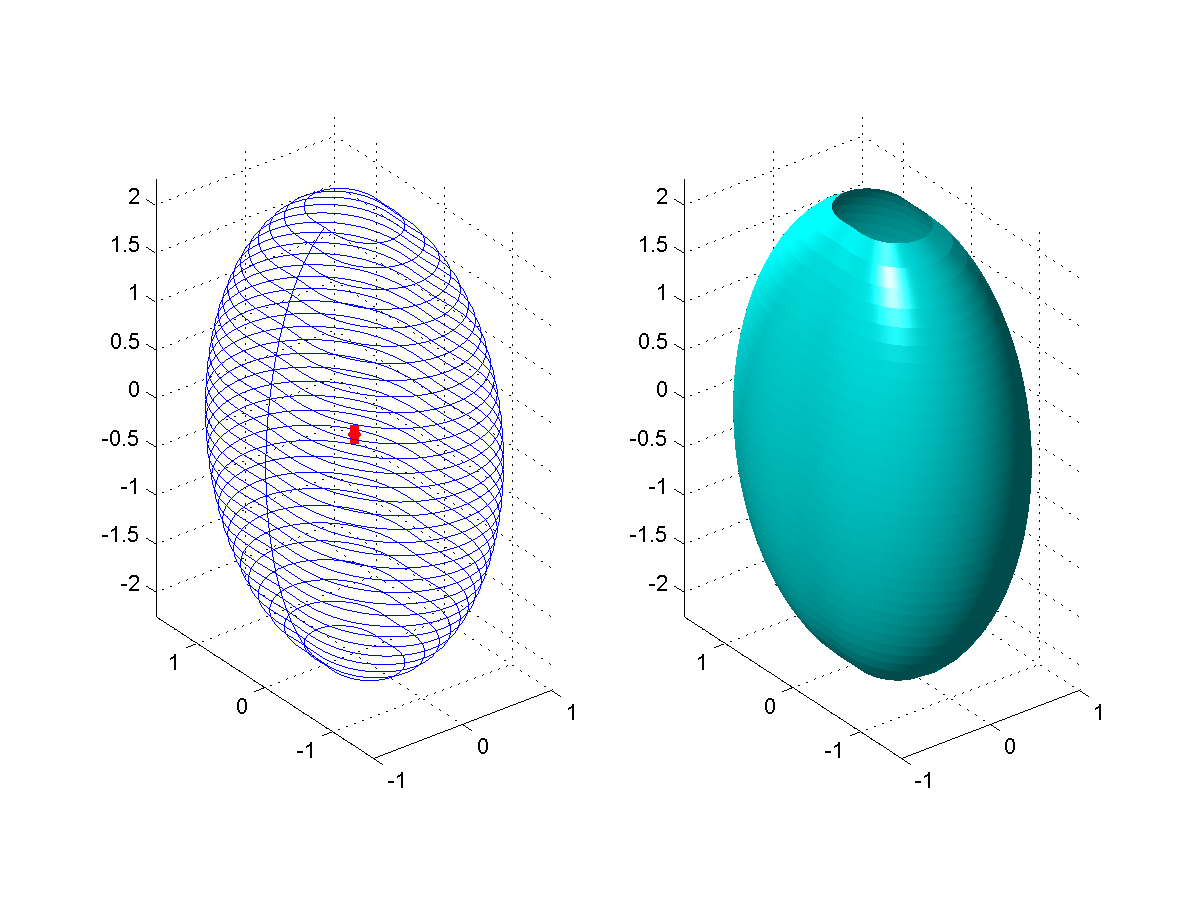}}
%%% sous figure 
\subfigure[\label{quelquesvolumes_N33_Q16} : $N=33$ et $Q=16$]
%{\epsfig{file=quelquesvolumes_N33_Q16.eps, width=5cm}}
{\includegraphics[width=5 cm]{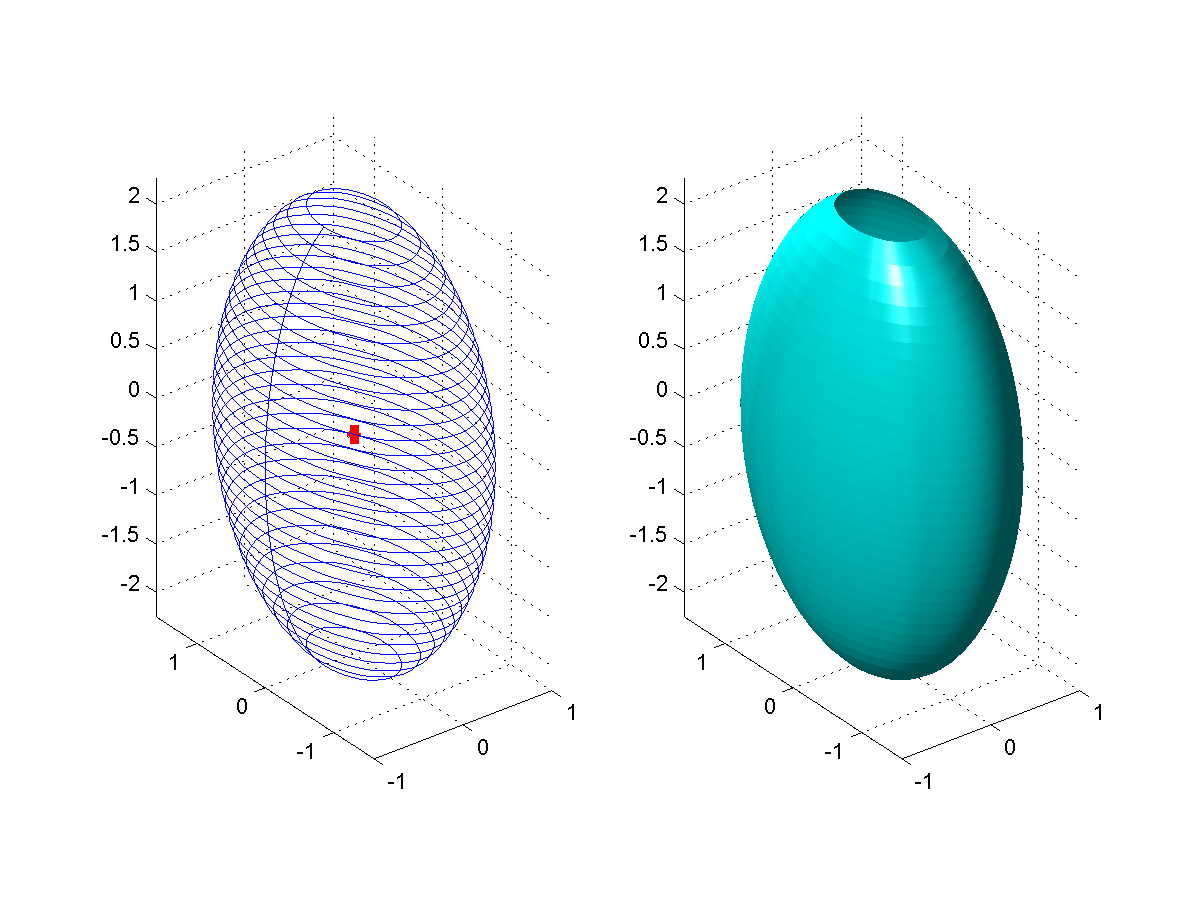}}
%%% sous figure 
\subfigure[\label{quelquesvolumes_N33_Q32} : $N=33$ et $Q=32$]
%{\epsfig{file=quelquesvolumes_N33_Q32.eps, width=5cm}}
{\includegraphics[width=5 cm]{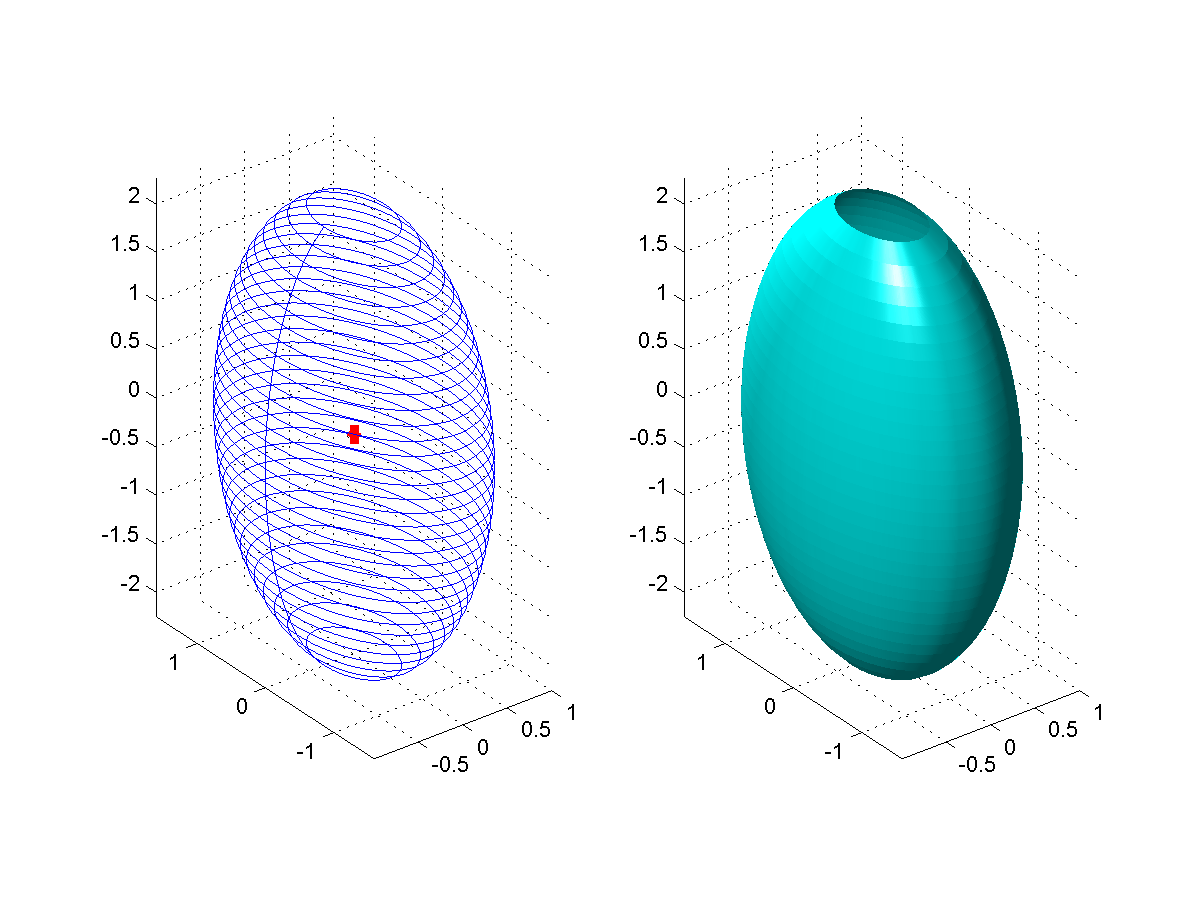}}
\caption{\label{valelifig10}Quelques volumes reconstitués de l'ellipsoïde en fonction des différentes valeurs de $N$ et de $Q$.}
\end{figure}

Voir la figure \ref{valelifig10} où quelques volumes ont été tracés. On voir sur cette figure que les volumes semblent dépendre surtout de $N$.

\begin{table}
\begin{center}
\begin{tabular}{lllll}
\hline
$N$ & & & & \\ \hline
$33$ &$108.26$ &$100.03$ &$99.87$ & $99.91$
\\ \hline
$17$ &$108.02$ &$99.81$ &$99.64$ & $99.69$
\\ \hline
$9$ &$107.27$ &$99.11$ &$98.95$ & $99.00$
\\ \hline
$5$ &$105.35$ &$97.33$ &$97.18$ & $97.22$
\\ \hline
$3$ &$101.58$ &$93.86$ &$93.71$ & $93.75$
\\ \hline
$Q$ &$4$ &$8$ &$16$ &$32$
 \\ \hline
\end{tabular}
\end{center}
\vspace{0.7 cm}
\caption{\label{valelitab10}Pourcentages du volume approché par rapport au volume réel.}
\end{table}

Voir aussi le tableau \ref{valelitab10} où ont été donnés les pourcentages du volume approché par rapport au volume réel.
On constate sur ce tableau que les pourcentages se rapprochent de 100, comme dans le tableau 1 de \cite{henderson1999}. La différence
est que les résulats sont plus précis et que, contrairement à Henderson, les pourcentages sont parfois supérieurs à 100.

\begin{table}
\begin{center}
\begin{tabular}{lllll}
\hline
$N$ & & & & \\ \hline
$33$ &$1.514 \, 10^{-16}$ &$4.139 \, 10^{-17}$ &$6.565 \, 10^{-17}$ & $1.209 \, 10^{-17}$
\\ \hline
$17$ &$7.751 \, 10^{-17}$ &$1.123 \, 10^{-17}$ &$3.395 \, 10^{-17}$ & $2.76 \, 10^{-17}$
\\ \hline
$9$ &$6.206 \, 10^{-17}$ &$2.786 \, 10^{-17}$ &$2.790 \, 10^{-17}$ & $1.839 \, 10^{-17}$
\\ \hline
$5$ &$8.746 \, 10^{-17}$ &$4.637 \, 10^{-17}$ &$2.841 \, 10^{-17}$ & $2.839 \, 10^{-17}$
\\ \hline
$3$ &$9.256 \, 10^{-17}$ &$2.762 \, 10^{-17}$ &$1.508 \, 10^{-17}$ & $1.499 \, 10^{-17}$
\\ \hline
$Q$ &$4$ &$8$ &$16$ &$32$
 \\ \hline
\end{tabular}
\end{center}
\vspace{0.7 cm}
\caption{\label{valelitab20}Erreurs commises sur la position du centre de gravité.}
\end{table}

\begin{table}
\begin{center}
\begin{tabular}{lllll}
\hline
$N$ & & & & \\ \hline
$33$ &$0.0964$ &$0.0064$ &$0.0050$ & $0.0032$
\\ \hline
$17$ &$0.0881$ &$0.0140$ &$0.0127$ & $0.0109$
\\ \hline
$9$ &$0.0682$ &$0.0378$ &$0.0364$ & $0.0346$
\\ \hline
$5$ &$0.0679$ &$0.0977$ &$0.0963$ & $0.0945$
\\ \hline
$3$ &$0.1390$ &$0.2103$ &$0.2087$ & $0.2069$
\\ \hline
$Q$ &$4$ &$8$ &$16$ &$32$
 \\ \hline
\end{tabular}
\end{center}
\vspace{0.7 cm}
\caption{\label{valelitab30}Erreurs commises sur les inerties.}
\end{table}

On a aussi affiché dans les tableaux 
\ref{valelitab20} et  
\ref{valelitab30},
les erreurs commises sur les évaluations du centre de gravité, très faibles, par raison de symétrie
mais aussi sur les inerties qui se rapprochent de zéro quand $N$ et $Q$ augmentent.

Enfin, pour $N=1000$ et $Q=1000$, on obtient les erreurs suivantes 
sur le volume, le centre de gravité et l'inertie :
\begin{align*}
&\varepsilon_V=9.98 \, 10^{-7},\\
&\varepsilon_{\text{CG}}=5.67 \, 10^{-17},\\
&\varepsilon_I=3.45 \, 10^{-6},
\end{align*}
ce qui est très faible.
%%%%%%%%%%%%%%%%%%%%%%%%%%%%%%%%%%%%%%%%%%%%%%%%%%%%%%%%%%%%%

%%%%%%%%%%%%%%%%%%%%%%%%%%%%%%%%%%%%%%%%%%%%%%%%%%%%%%%%%%%%%
\subsection{Exemple du Pogona}
\label{simulation_pogona}

%%%%%%%%%%%%%%%%%%%%%%%%%%%%%%%%%%%%%%%%%%%%%%%%%%%%%%%%%%%%%
%\input{exemple_Pogona}
% fichier tex crée par MaTeXBuild02 le 23-Jun-2014 11:23:15
% à compiler avec MaTeXBuild02('exemple_Pogona',0)
% ATTENTION, long !!!

En choisissant un centre de section fixe et nul (voir remarque \ref{rem00}), on a pu 
utiliser les 4 méthodes. 
Les calculs des données morphologiques (volume, position du centre de masse et matrices d'inerties données
par \eqref{eq300},\eqref{eq310} et \eqref{eq320}, notés avec des indices X pour X dans $\{\text{A},\text{B},\text{C},\text{D}\}$ 
en fonction des quatre méthodes présentées)
sont présentées ci-dessous, pour les différentes méthodes utilisées.
On obtient pour le volume
\begin{subequations}
\label{eq300totnumNEW}
\begin{align}
&V_{\text{A}}=0.3129,\\
&V_{\text{B}}=0.3138,\\
&V_{\text{C}}=0.3126,\\
&V_{\text{D}}=0.3028,
\end{align}
\end{subequations}
pour la position du centre de masse
\begin{subequations}
\label{eq310totnumNEW}
\begin{align}
\label{eq310anumNEW}
&G_{\text{A}}=
\begin{pmatrix}
0.0034\\
0.0327\\
0.6473
\end{pmatrix}
,\\
\label{eq310bnumNEW}
&G_{\text{B}}=
\begin{pmatrix}
0.0037\\
0.0387\\
0.6473
\end{pmatrix}
,\\
\label{eq310cnumNEW}
&G_{\text{C}}=
\begin{pmatrix}
0.0034\\
0.0318\\
0.6476
\end{pmatrix}
,\\
\label{eq310dnumNEW}
&G_{\text{D}}=
\begin{pmatrix}
0.0029\\
0.0325\\
0.6397
\end{pmatrix}
,
\end{align}
\end{subequations}
et pour les matrices d'inerties
\begin{subequations}
\label{eq320totnumNEW}
\begin{align}
\label{eq320anumNEW}
&I_{\text{A}}=
\begin{pmatrix}
0.0239 &0.0000 & -0.0004 \\
0.0000 &0.0325 & 0.0008 \\
-0.0004 &0.0008 & 0.0211 \\
\end{pmatrix}
,\\
\label{eq320bnumNEW}
&I_{\text{B}}=
\begin{pmatrix}
0.0240 &0.0000 & -0.0005 \\
0.0000 &0.0327 & 0.0008 \\
-0.0005 &0.0008 & 0.0212 \\
\end{pmatrix}
,\\
\label{eq320cnumNEW}
&I_{\text{C}}=
\begin{pmatrix}
0.0239 &0.0000 & -0.0004 \\
0.0000 &0.0325 & 0.0007 \\
-0.0004 &0.0007 & 0.0210 \\
\end{pmatrix}
,\\
\label{eq320dnumNEW}
&I_{\text{D}}=
\begin{pmatrix}
0.0229 &0.0000 & -0.0004 \\
0.0000 &0.0297 & 0.0006 \\
-0.0004 &0.0006 & 0.0194 \\
\end{pmatrix}
.
\end{align}
\end{subequations}
On observe donc des résultats très proches pour les quatre méthodes.

\begin{figure}[h] 
\begin{center} 
\includegraphics[width=15 cm]{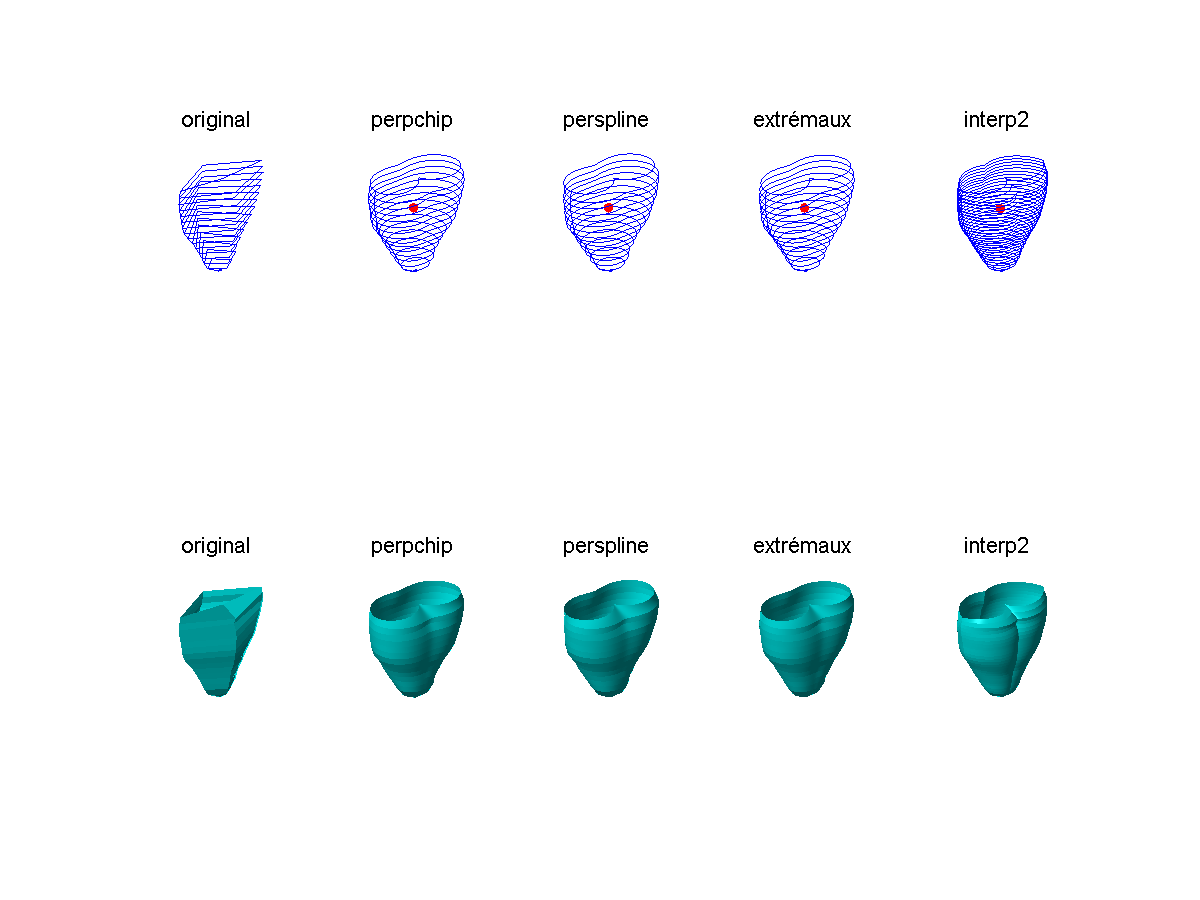}
\end{center} 
\caption{\label{exemple_Pogona}Exemple de simulation.} 
\end{figure}

On présente en figure \ref{exemple_Pogona} les résultats des quatre méthodes appliquées au Pogona de la figure \ref{fig01tot}.

En choisissant maintenant, un centre de section égal au centre de masse des point expérimentaux (voir remarque \ref{rem00}), on a pu 
utiliser les 3 méthodes.

Les calculs des données morphologiques 
sont présentées ci-dessous, pour les trois méthodes utilisées.
On obtient pour le volume
\begin{subequations}
\label{eq300totnum}
\begin{align}
\label{eq300anum}
&V_{\text{A}}=0.3123,\\
&V_{\text{B}}=0.3137,\\
&V_{\text{C}}=0.3122,
\end{align}
\end{subequations}
pour la position du centre de masse
\begin{subequations}
\label{eq310totnum}
\begin{align}
\label{eq310anum}
&G_{\text{A}}=
\begin{pmatrix}
0.0037\\
0.0379\\
0.6473
\end{pmatrix}
,\\
\label{eq310bnum}
&G_{\text{B}}=
\begin{pmatrix}
0.0035\\
0.0428\\
0.6472
\end{pmatrix}
,\\
\label{eq310cnum}
&G_{\text{C}}=
\begin{pmatrix}
0.0037\\
0.0378\\
0.6474
\end{pmatrix}
,
\end{align}
\end{subequations}
et pour les matrices d'inerties
\begin{subequations}
\label{eq320totnum}
\begin{align}
\label{eq320anum}
&I_{\text{A}}=
\begin{pmatrix}
0.0238 &0.0000 & -0.0005 \\
0.0000 &0.0325 & 0.0007 \\
-0.0005 &0.0007 & 0.0211 \\
\end{pmatrix}
,\\
\label{eq320bnum}
&I_{\text{B}}=
\begin{pmatrix}
0.0240 &0.0000 & -0.0005 \\
0.0000 &0.0328 & 0.0007 \\
-0.0005 &0.0007 & 0.0213 \\
\end{pmatrix}
,\\
\label{eq320cnum}
&I_{\text{C}}=
\begin{pmatrix}
0.0238 &0.0000 & -0.0005 \\
0.0000 &0.0325 & 0.0007 \\
-0.0005 &0.0007 & 0.0210 \\
\end{pmatrix}
.
\end{align}
\end{subequations}
On observe donc des résultats très proches pour les trois méthodes.
%%%%%%%%%%%%%%%%%%%%%%%%%%%%%%%%%%%%%%%%%%%%%%%%%%%%%%%%%%%%%

%%%%%%%%%%%%%%%%%%%%%%%%%%%%%%%%%%%%%%%%%%%%%%%%%%%%%%%%%%%%%
\subsection{Comparaison de quelques méthodes}
\label{simulation_compare}

%%%%%%%%%%%%%%%%%%%%%%%%%%%%%%%%%%%%%%%%%%%%%%%%%%%%%%%%%%%%%
%\input{comparaison_quelques_methodes}
% fichier tex crée par MaTeXBuild02 le 23-Jun-2014 11:18:09
% à compiler avec MaTeXBuild02('comparaison_quelques_methodes',0)
% ATTENTION, très long !!!

Conformément à ce que l'on a présenté en remarque \ref{rem00}
et en section \ref{gencylin}, 
on peut faire le calcul avec plusieurs possibilités : 
\begin{itemize}
\item
en utilisant comme centre de section, le barycentre
des points mesurés  ou un point fixe (choisi de coordonnées nulles) ;
\item
en calculant l'intégrale en $\theta$ de façon exacte ou approchée.
\end{itemize}

En choisissant un calcul approché de l'intégrale en $\theta$, 
on a comparé le calcul avec calcul avec barycentre (résultats obtenus dans 
\eqref{eq300totnum},
\eqref{eq310totnum}
et
\eqref{eq320totnum})
 et sans barycentre.
L'écart obtenu entre les deux méthodes
vaut 
\begin{itemize}
\item[$\bullet$]
pour le volume 
$5.527 \, 10^{-4}$ ; 
\item[$\bullet$]
pour la position du centre de gravité
$5.985 \, 10^{-3}$ ; 
\item[$\bullet$]
pour la matrice d'inertie 
$1.349 \, 10^{-4}$.
\end{itemize}
Ces écarts étant faibles, on utilisera donc par la suite la méthode avec barycentre,
ce qui autorise aussi un éventule décalage  des sections par rapport à un axe de référence.

En choisissant le calcul avec calcul avec barycentre,
on a aussi comparé le  calcul approché de l'intégrale en $\theta$
avec $1001$ points 
(résultats obtenus dans 
\eqref{eq300totnum},
\eqref{eq310totnum}
et
\eqref{eq320totnum})
avec le calcul exact de cette intégrale.
On obtient un écart total (pour le volume, centre de gravité et inerties) égal à 
$7.197 \, 10^{-11}$
et un rapport de temps d'exécution (exact/approché), égal à 
$1.652867$.
De même, si on choisit le calcul avec origine nulle de section, 
on obtient un écart total  égal à 
$1.286 \, 10^{-12}$
et un rapport de temps d'exécution  égal à 
$5.230257$.
On conservera donc pour la suite le  calcul approché de l'intégrale en $\theta$
avec $1001$ points.

%%%%%%%%%%%%%%%%%%%%%%%%%%%%%%%%%%%%%%%%%%%%%%%%%%%%%%%%%%%%%

%%%%%%%%%%%%%%%%%%%%%%%%%%%%%%%%%%%%%%%%%%%%%%%%%%%%%%%%%%%%%
\subsection{Exemple du lézard}
\label{simulation_lezard}

%%%%%%%%%%%%%%%%%%%%%%%%%%%%%%%%%%%%%%%%%%%%%%%%%%%%%%%%%%%%%
%\input{exemple_lezard}
% fichier tex crée par MaTeXBuild02 le 23-Jun-2014 11:51:49
% à compiler avec clear all;MaTeXBuild02('exemple_lezard',0)

\begin{figure}[h] 
\begin{center} 
\includegraphics[width=13 cm]{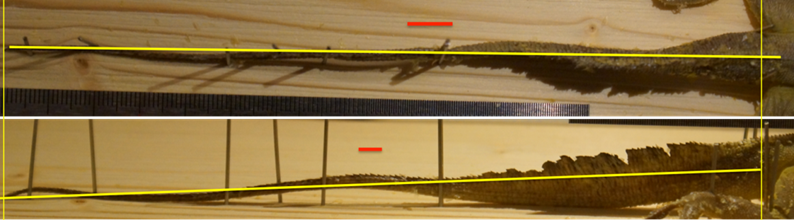}
\end{center} 
\caption{\label{exemple_lezard_photo_queue}Photos du membre considéré (queue).} 
\end{figure}

\begin{figure}[h] 
\begin{center} 
\includegraphics[width=12 cm]{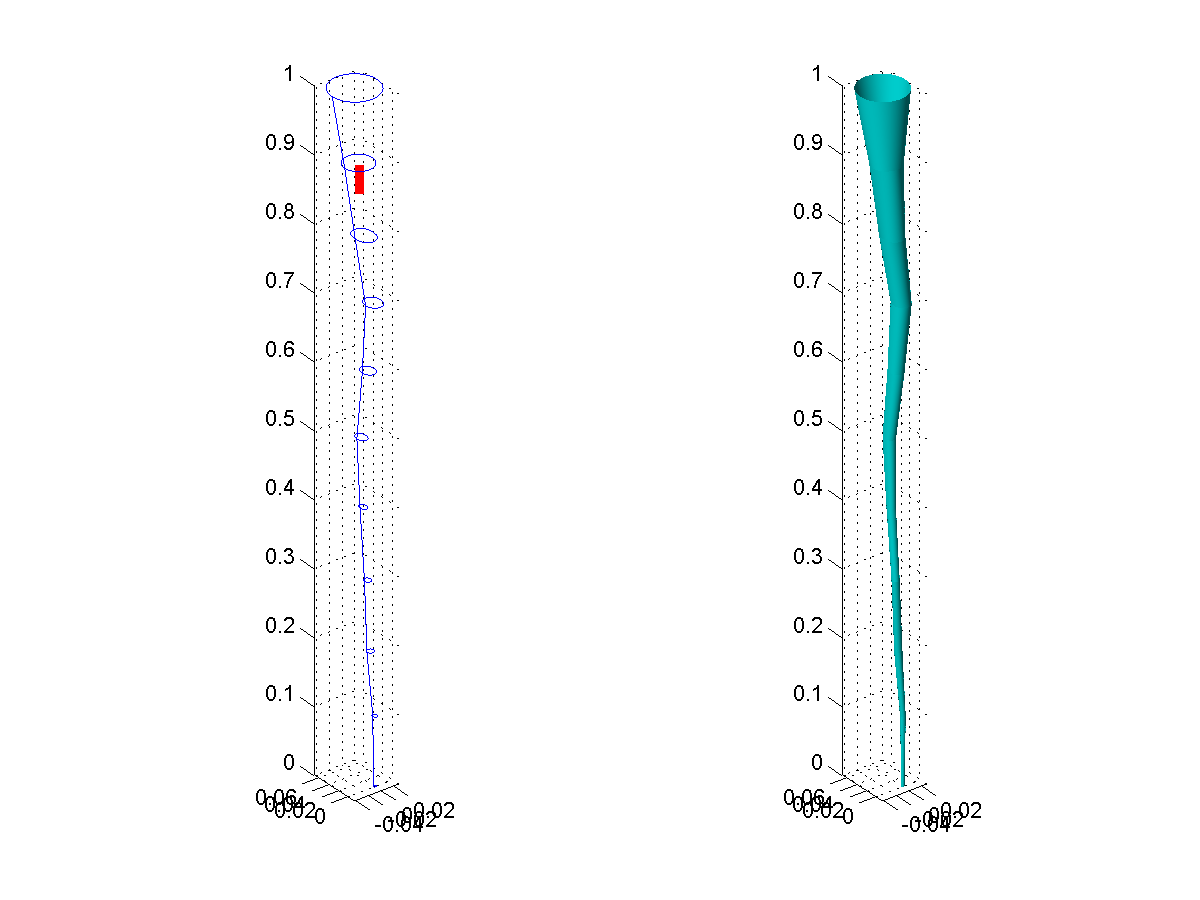}
\end{center} 
\caption{\label{exemple_lezard_queue}Exemple de simulation sur le lézard (queue).} 
\end{figure}

On présente en figure \ref{exemple_lezard_queue} les résultats de la méthode conservée (c'est-à-dire la A) pour le lézard
(\textit{Anolis ferreus} :  queue) dont le membre étudié est représenté en figure \ref{exemple_lezard_photo_queue}.

On obtient pour le volume
\begin{subequations}
\label{exemple_lezard_res_queue}
\begin{equation}
V=0.0008,
\end{equation}
pour la position du centre de masse
\begin{equation}
G=
\begin{pmatrix}
-0.0086\\
0.0215\\
0.8721
\end{pmatrix}
.
\end{equation}
et pour les matrices d'inerties
\begin{equation}
I_=
1.0 \, 10^{-5}\times
\begin{pmatrix}
2.7939 &-0.0001 & 0.0122 \\
-0.0001 &2.7833 & -0.1308 \\
0.0122 &-0.1308 & 0.0429 \\
\end{pmatrix}
\end{equation}
\end{subequations}

%%% Copier-colle de ce qui précéde sauf XX=

\begin{figure}[h] 
\begin{center} 
\includegraphics[width=13 cm]{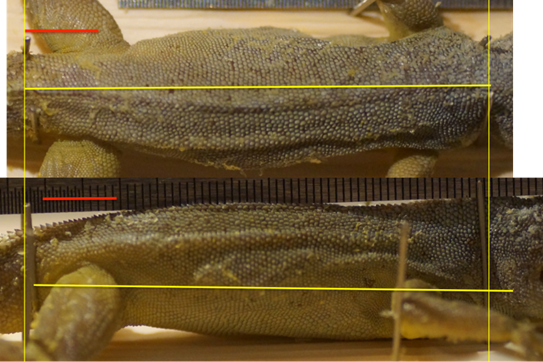}
\end{center} 
\caption{\label{exemple_lezard_photo_tronc}Photos du membre considéré (tronc).} 
\end{figure}

\begin{figure}[h] 
\begin{center} 
\includegraphics[width=12 cm]{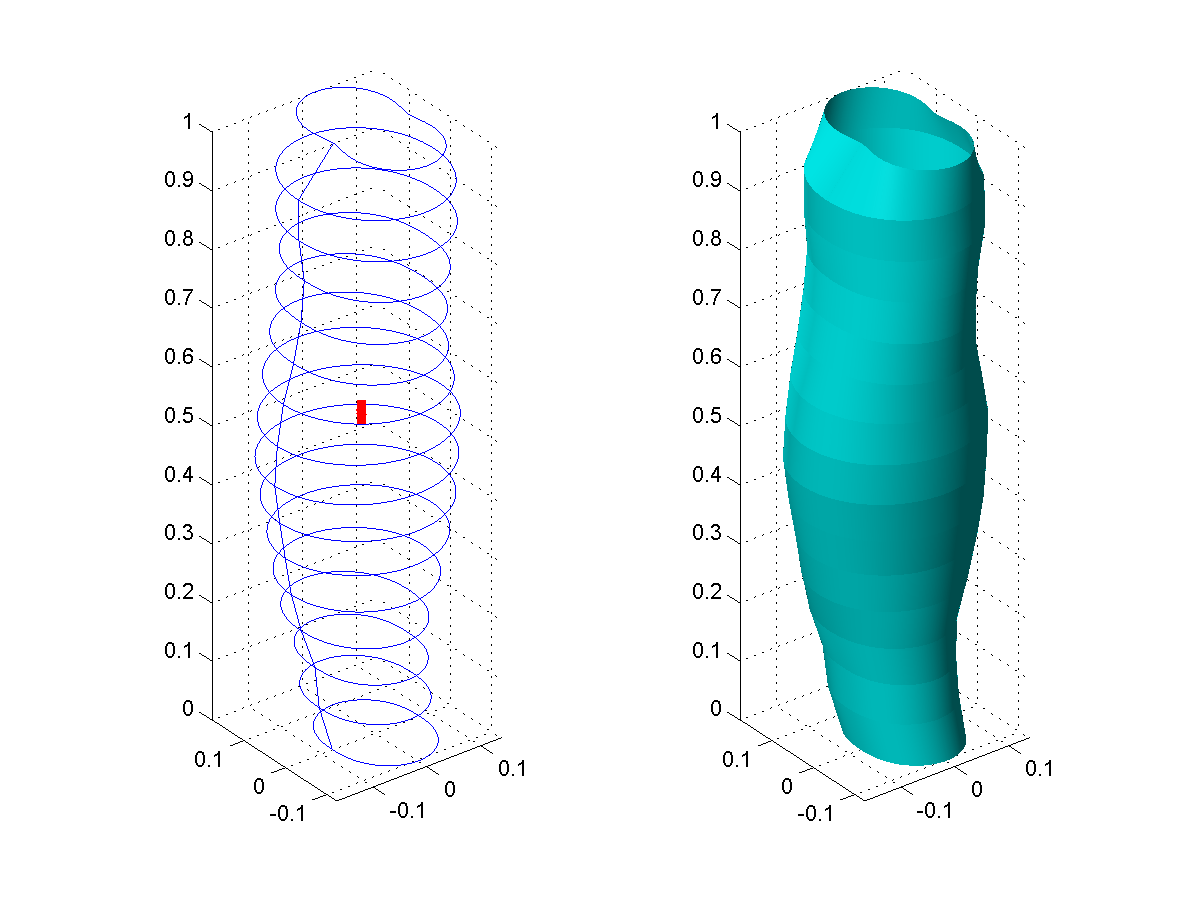}
\end{center} 
\caption{\label{exemple_lezard_tronc}Exemple de simulation sur le lézard (tronc).} 
\end{figure}

On présente en figure \ref{exemple_lezard_tronc} les résultats de la méthode conservée (c'est-à-dire la A) pour le lézard
(\textit{Anolis ferreus} :  tronc) dont le membre étudié est représenté en figure \ref{exemple_lezard_photo_tronc}.

On obtient pour le volume
\begin{subequations}
\label{exemple_lezard_res_tronc}
\begin{equation}
V=0.0536,
\end{equation}
pour la position du centre de masse
\begin{equation}
G=
\begin{pmatrix}
-0.0077\\
0.0251\\
0.5365
\end{pmatrix}
.
\end{equation}
et pour les matrices d'inerties
\begin{equation}
I_=
1.0 \, 10^{-5}\times
\begin{pmatrix}
428.7112 &-0.1112 & -6.8173 \\
-0.1112 &424.7810 & -3.3236 \\
-6.8173 &-3.3236 & 47.3539 \\
\end{pmatrix}
\end{equation}
\end{subequations}

%%% Copier-colle de ce qui précéde sauf XX=
%%%%%%%%%%%%%%%%%%%%%%%%%%%%%%%%%%%%%%%%%%%%%%%%%%%%%%%%%%%%%

%%%%%%%%%%%%%%%%%%%%%%%%%%%%%%%%%%%%%%%%%%%%%%%%%%%%%%%%%%%%%
\subsection{Comparaison avec le cylindre}
\label{simulation_compare_cylindre}

%%%%%%%%%%%%%%%%%%%%%%%%%%%%%%%%%%%%%%%%%%%%%%%%%%%%%%%%%%%%%
%\input{compare_cylindre}
% fichier tex crée par MaTeXBuild02 le 23-Jun-2014 11:23:53
% à compiler avec MaTeXBuild02('compare_cylindre',0)

\begin{remark}
Pour le tronc du lézard, proche d'un cylindre, on a aussi calculé les données morphométriques
en considérant un cylindre moyen. Ce cylindre est de hauteur $h$ (sur l'axe $z$) et de rayon $R$.
On vérifie que
\begin{equation*}
I_x=\iiint _{\mathcal{D}} x^2dV=
\iiint _{\mathcal{D}} r^2\cos^2\theta r  drd\theta dz=
\int_{-h/2}^{h/2} dz \int_{-\pi}^\pi \cos^2d\theta\int_0^R r^3dr,
\end{equation*}
et donc 
\begin{equation*}
I_x=\frac{h\pi R ^4}{4}.
\end{equation*}
De même, 
\begin{equation*}
I_y=\iiint _{\mathcal{D}} y^2dV=I_x.
\end{equation*}
Enfin, 
\begin{equation*}
I_z=\iiint _{\mathcal{D}} z^2dV=
\iiint _{\mathcal{D}} z^2 rdrd\theta dz=
\int_{-h/2}^{h/2} z^2 dz 
\iint _{\substack{\theta\in [-\pi,\pi]\\r\in [0,R]}}r  drd\theta=
2\int_{0}^{h/2} z^2 dz \times \pi R^2
=\frac{2}{3}{\left(\frac{h}{2}\right)}^3 \times \pi R^2
\end{equation*}
et donc 
\begin{equation*}
I_z=\frac{h^3\pi R ^2}{12}.
\end{equation*}
Par symétrie (ou imparité) les termes non diagonaux de la matrice d'inertie sont nuls et on a 
\begin{equation*}
I=
\mu
\begin{pmatrix}
\iiint _\mathcal{D}  y^2+z^2dV&0 & 0 \\
0  & \iiint _\mathcal{D} x^2+z^2  dV & 0 \\
0 & 0 &  \iiint _\mathcal{D} x^2+y^2 
\end{pmatrix},
\end{equation*}
et donc 
\begin{equation*}
I=
\mu
\frac{h\pi R}{4}
\begin{pmatrix}
R^3+\frac{h^2R}{3}&0 & 0 \\
0 & R^3+\frac{h^2R}{3} & 0 \\
0 & 0 & 2R^3 
\end{pmatrix}.
\end{equation*}

Les résultats sont du même ordre que ceux obtenus précédemment
(voir \eqref{exemple_lezard_res_tronc}).
On obtient pour le volume
\begin{equation*}
V=0.0477,
\end{equation*}
pour la position du centre de masse
\begin{equation*}
G=
\begin{pmatrix}
-0.0000\\
0.0059\\
0.5000
\end{pmatrix}
.
\end{equation*}
et pour les matrices d'inerties
\begin{equation*}
I_=
1.0 \, 10^{-5}\times
\begin{pmatrix}
415.2233 &0.0000 & 0.0000 \\
0.0000 &415.2233 & 0.0000 \\
0.0000 &0.0000 & 36.1485 \\
\end{pmatrix}
\end{equation*}
\end{remark}
%%%%%%%%%%%%%%%%%%%%%%%%%%%%%%%%%%%%%%%%%%%%%%%%%%%%%%%%%%%%%

%%%%%%%%%%%%%%%%%%%%%%%%%%%%%%%%%%%%%%%%%%%%%%%%%%%%%%%%%%%%%
% Figure obtenue avec 
%load Pogona;
%principalA(verts,[],0,0,1);
%stopper à la section d'altitude 0.2667 et sauvegarder l'image en eps dans section_cote0p2667.eps
%%%%%%%%%%%%%%%%%%%%%%%%%%%%%%%%%%%%%%%%%%%%%%%%%%%%%%%%%%%%%

\begin{figure}[h] 
\begin{center} 
\includegraphics[width=14 cm]{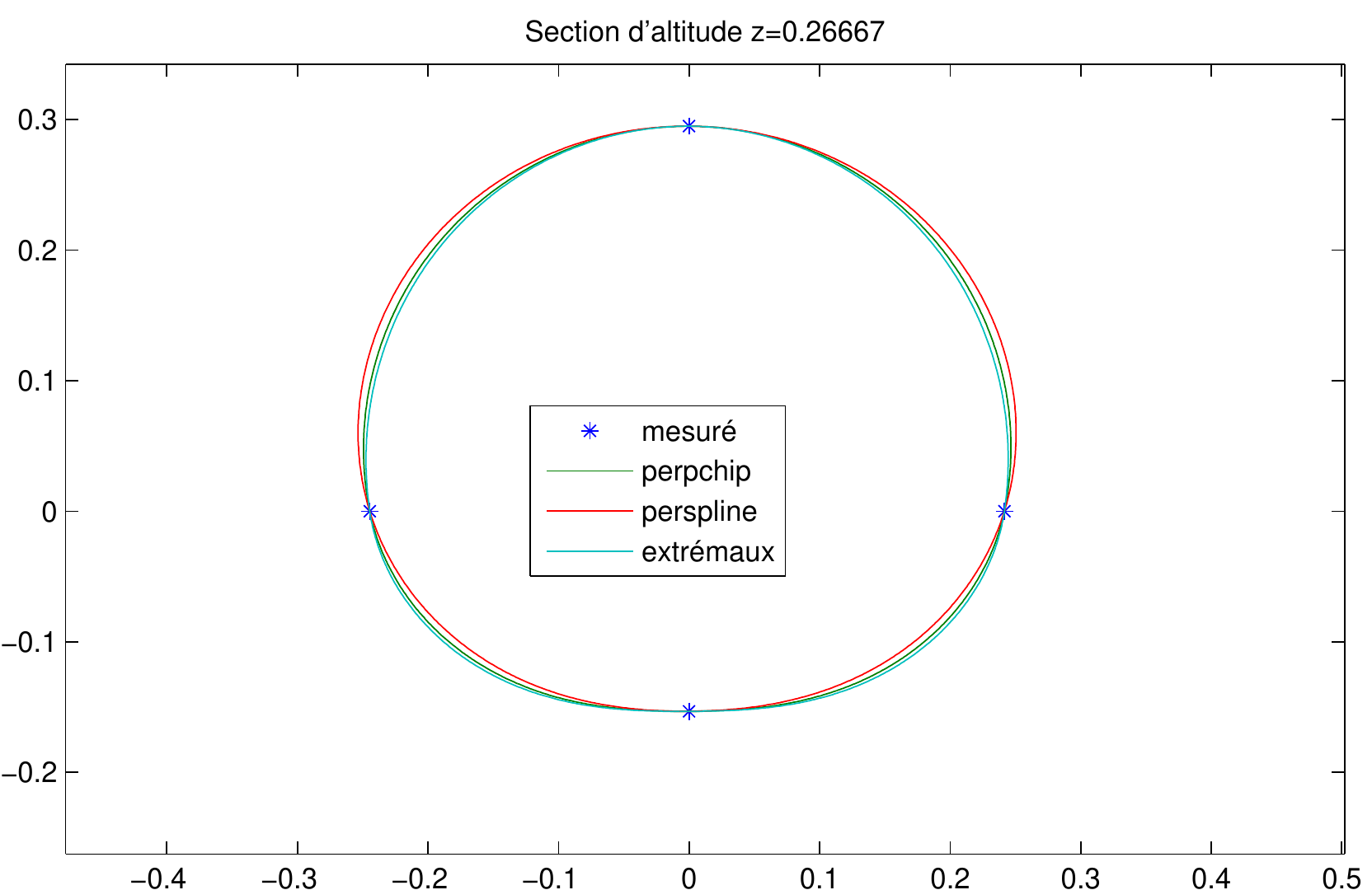}
\end{center} 
\caption{\label{fig100}La frontière $\partial \mathcal{D}(z_0)$.} 
\end{figure}

Sur la figure \ref{fig100}, a été tracée la frontière $\partial \mathcal{D}(z_0)$
correspondant à $z_0=0.2667$. On constate sur cette figure que les trois  méthodes donnent des résultats très proches.

%%%%%%%%%%%%%%%%%%%%%%%%%%%%%%%%%%%%%%%%%%%%%%%%%%%%%%%%%%%%%
% Figure obtenue avec 
%load Pogona;
%principalA(verts,centresect,0,0,1);
%stopper à la section d'altitude 0.2667 et sauvegarder l'image en eps dans section_cote0p2667bis.eps
%%%%%%%%%%%%%%%%%%%%%%%%%%%%%%%%%%%%%%%%%%%%%%%%%%%%%%%%%%%%%

\begin{figure}[h] 
\begin{center} 
\includegraphics[width=14 cm]{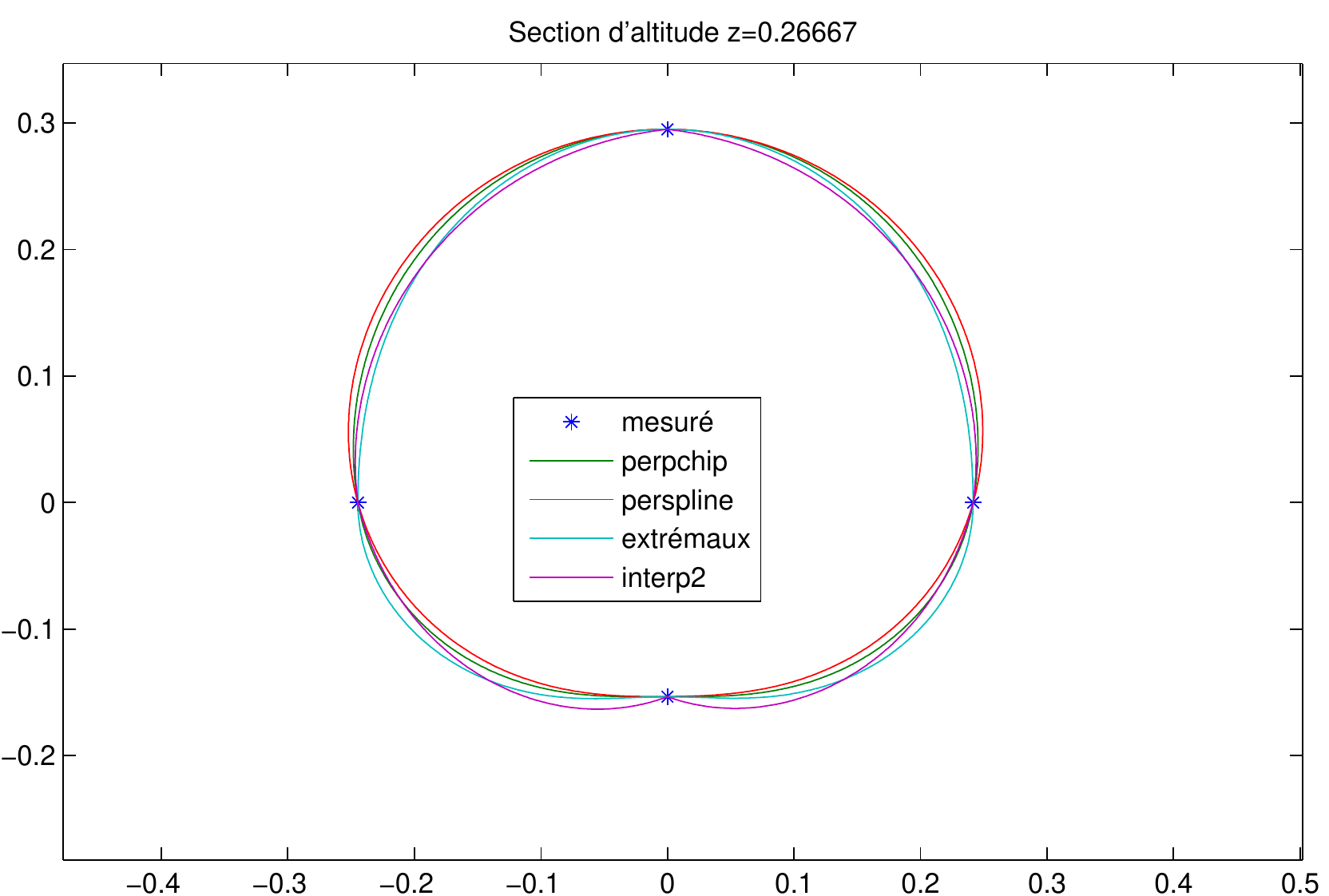}
\end{center} 
\caption{\label{fig100bis}La frontière $\partial \mathcal{D}(z_0)$.} 
\end{figure}

Pour comparer aussi avec la méthode D, on a tracé en \ref{fig100bis}, a été tracée la frontière $\partial \mathcal{D}(z_0)$
correspondant à $z_0=0.2667$ avec centre de section nul. On constate sur cette figure que les quatre méthodes donnent des résultats très proches. 
Comme annoncé,
la frontière correspondant à la méthode D présente un point anguleux.
Cette image conforte l'idée que chaque section est proche d'un cercle et, qu'en coordonnées polaires,
$r(\theta)$, ne varie que peu en $\theta$.

%%%%%%%%%%%%%%%%%%%%%%%%%%%%%%%%%%%%%%%%%%%%%%%%%%%%%%%%%%%%%
% Figure obtenue avec 
%load Pogona;
%principalA(verts,centresect,1);
%puis stoper et sauvegarder la première image en eps dans figinterP2.eps
%%%%%%%%%%%%%%%%%%%%%%%%%%%%%%%%%%%%%%%%%%%%%%%%%%%%%%%%%%%%%

\begin{figure}[h] 
\begin{center} 
\includegraphics[width=10 cm]{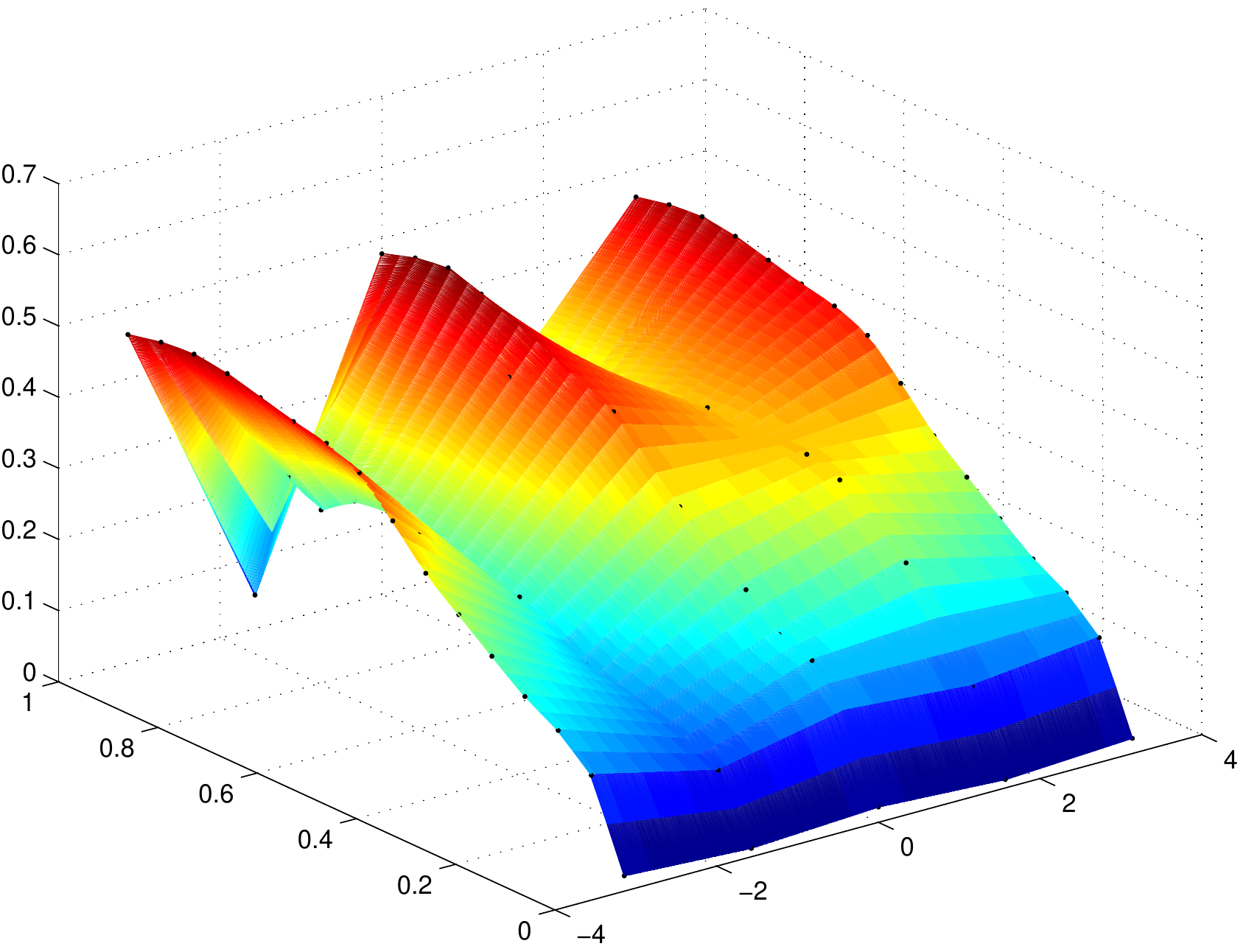}
\end{center} 
\caption{\label{fig110}La surface $(\theta,z)\mapsto \rho(\theta,z)$ de la méthode D et les points expérimentaux.} 
\end{figure}

Sur la figure \ref{fig110}, a été tracée la surface $(\theta,z)\mapsto \rho(\theta,z)$ de la méthode D (avec centre de section nul) et les points expérimentaux.

%%%%%%%%%%%%%%%%%%%%%%%%%%%%%%%%%%%%%%%%%%%%%%%%%%%%%%%%%%%%%
%%%%%%%%%%%%%%%%%%%%%%%%%%%%%%%%%%%%%%%%%%%%%%%%%%%%%%%%%%%%%
\section{Comparaison avec les ellipses de Henderson}
\label{comaparaisonavechenderson}

Puisque notre méthode correspond à la méthode de Henderson quand $Q=4$, montrons que les résultats que l'on obtient
sont proches de ceux qu'auraient obtenus Henderson.

\begin{figure}
% \psfrag{a}{$a$}
% \psfrag{b}{$b$}
% \psfrag{ma}{$-a$}
% \psfrag{mb}{$-b$}
% \psfrag{r1}[][l]{$r_1(\theta)$}
% \psfrag{r2}{$r_2(\theta)$}
% \psfrag{th}{$\theta$}
\centering
%%% sous figure 1
\subfigure[\label{explication_compare_ellipse}Utilisation de la fonction \texttt{perpchip} seule]
%{\epsfig{file=dudu.eps, width=4.5cm}}
{\includegraphics[width=4.5 cm]{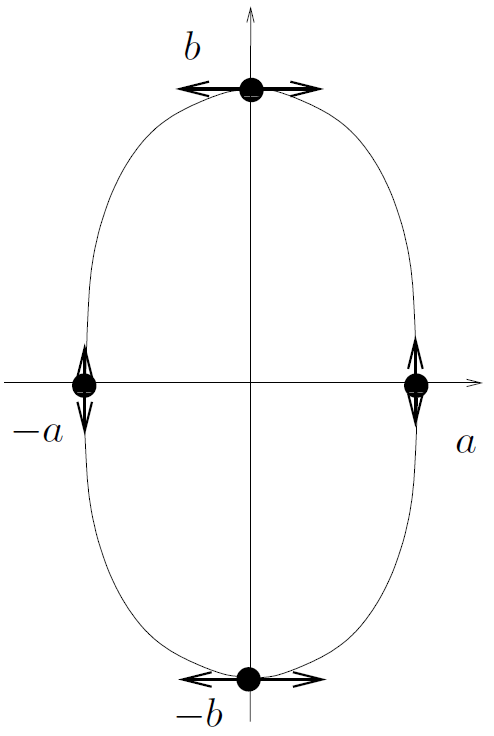}}
\qquad
%%% sous figure 2
\subfigure[\label{explication_compare_ellipse2}Utilisation de la fonction \texttt{perpchip} seule et ellipse de Henderson]
%{\epsfig{file=dudu2.eps, width=4.5cm}}
{\includegraphics[width=4.5 cm]{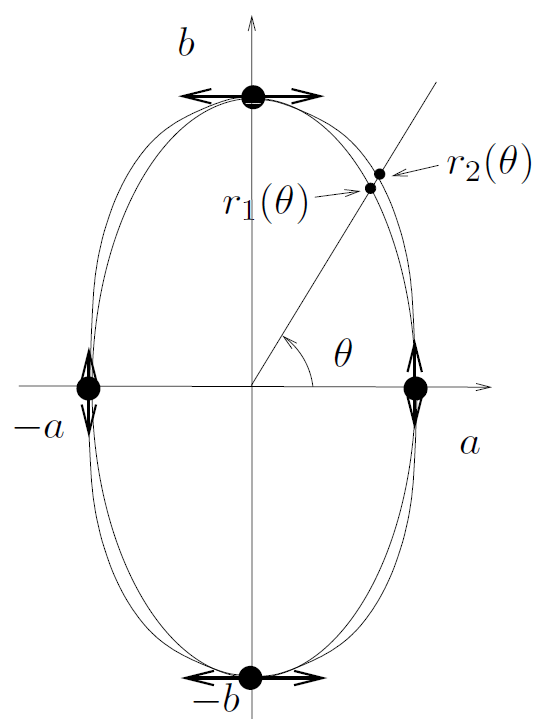}}
\caption{\label{explication_compare_ellipsetot}Utilisation de la fonction \texttt{perpchip}
dans le cas de 4 points symétriques et comparaison avec l'ellipse de Henderson.}
\end{figure}

Pour cela, il suffit de remarquer 
que l'utilisation de la fonction \path|perpchip|
avec 4 points symétriques donne une fonction $r(\theta)$ dont les dérivées sont nulles
aux valeurs de $\theta$ multiples de $\pi/2$. Voir la figure \ref{explication_compare_ellipse2}.

Pour comparer les deux solutions obtenus, il suffit de calculer les deux fonctions
$r_1$ et $r_2$ correspondant aux coordonnées polaires respectivement donnés pour la solution avec \path|perpchip| et avec l'ellispe.

On peut par symétrie, supposer que $\theta$ décrit $[0,\pi/2]$. La fonction $r_1$ donnée par 
\path|perpchip|
et donc l'unique fonction $f$ de degré 3 sur $[0,\pi/2]$ telle que  
\begin{align*}
&f(0)=a,\\
&f'(0)=0,\\
&f(\pi/2)=b,\\
&f'(\pi/2)=0.
\end{align*}
Grâce à la théorie de l'interpolation (voir par exemple \cite{jbjnmdunod03}), on montre que nécessairement, on a 
\begin{equation}
\label{comelleq01}
r_1(\theta)=a+\frac{4(b-a)}{\pi^2}\theta^2-\frac{16(b-a)}{\pi^3}\theta^2(\theta-\pi/2)
\end{equation}
L'ellipse de demi-axes $a$ et $b$ est d'équation
\begin{equation}
\label{comelleq10}
\frac{x^2}{a^2}
+
\frac{y^2}{b^2}=1.
\end{equation}
Si on pose $x=r_2\cos\theta$ et $y=r_2\sin\theta$, on a donc
\begin{equation}
\label{comelleq20}
r_2(\theta)=
{\left(\frac{\cos^2\theta}{a^2}
+
\frac{\sin^2\theta}{b^2}\right)}^{-1/2}
\end{equation}
Pour caluler l'écart relatif entre les deux courbes, il suffit de calculer l'écart maximun entre $r_1$ et $r_2$.
puisque $r_1(0)=r_2(0)$ et $r_1(\pi/2)=r_2(\pi/2)$,
Il suffit de donc de résoudre $r'_1(\theta)-r'_2(\theta)=0$
sur $]0,\pi/2[$,
soit encore l'équation 
\begin{equation}
\label{comelleq21}
\frac{8(b-a)}{\pi^2} \theta^2-\frac{16(b-a)}{\pi^3}\left(2\theta(\theta-\pi/2)+\theta^2\right)-
\frac{1}{2}\left(\frac{1}{a^2}-\frac{1}{b^2}\right){\left(\frac{\cos^2\theta}{a^2}
+
\frac{\sin^2\theta}{b^2}\right)}^{-3/2}\sin2\theta=0
\end{equation}
que l'on résoud par un solveur et on en déduit l'écart maximum.

On peut aussi comparer l'aire entre les deux courbes.
Pour l'ellipse  on a, en se ramenant à un quart de l'aire totale 
\begin{equation}
\label{comelleq30}
\mathcal{A}=\frac{1}{4}\pi ab.
\end{equation}
Pour l'ellipse donnée par \eqref{comelleq01}, on a, en passant en coordonnées polaires  
\begin{equation*}
\mathcal{A}'=
\int_{\theta=0}^{\pi/2} \int_{r=0}^{r_1(\theta)} rdrd\theta.
\end{equation*}
On obtient donc 
\begin{equation*}
\mathcal{A}'=\frac{1}{2}
\int_{\theta=0}^{\pi/2} {\left[ a+\frac{4(b-a)}{\pi^2}\theta^2-\frac{16(b-a)}{\pi^3}\theta^2(\theta-\pi/2)\right]}^2d\theta.
\end{equation*}
On détermine cette intégrale en symbolique par exemple et on obtient 
\begin{equation}
\label{comelleq40}
\mathcal{A}'=\frac{\pi}{140}(13a^2 + 9ab + 13b^2)
\end{equation}
On n'a plus qu'à calculer 
\begin{equation}
\label{comelleq50}
\Delta S=
\frac{\left|\pi a b -4 \mathcal{A}'\right|}{\pi a b}.
\end{equation}

On vérifie que si $a=b$, les deux erreurs sont nulles.

%%%%%%%%%%%%%%%%%%%%%%%%%%%%%%%%%%%%%%%%%%%%%%%%%%%%%%%%%%%%%
%\input{simulation_compare_ellipse}
% fichier tex crée par MaTeXBuild02 le 23-Jun-2014 11:24:12
% à compiler avec MaTeXBuild02('simulation_compare_ellipse',0)
% très très long !!

\begin{figure}[h] 
\begin{center} 
\includegraphics[width=7 cm]{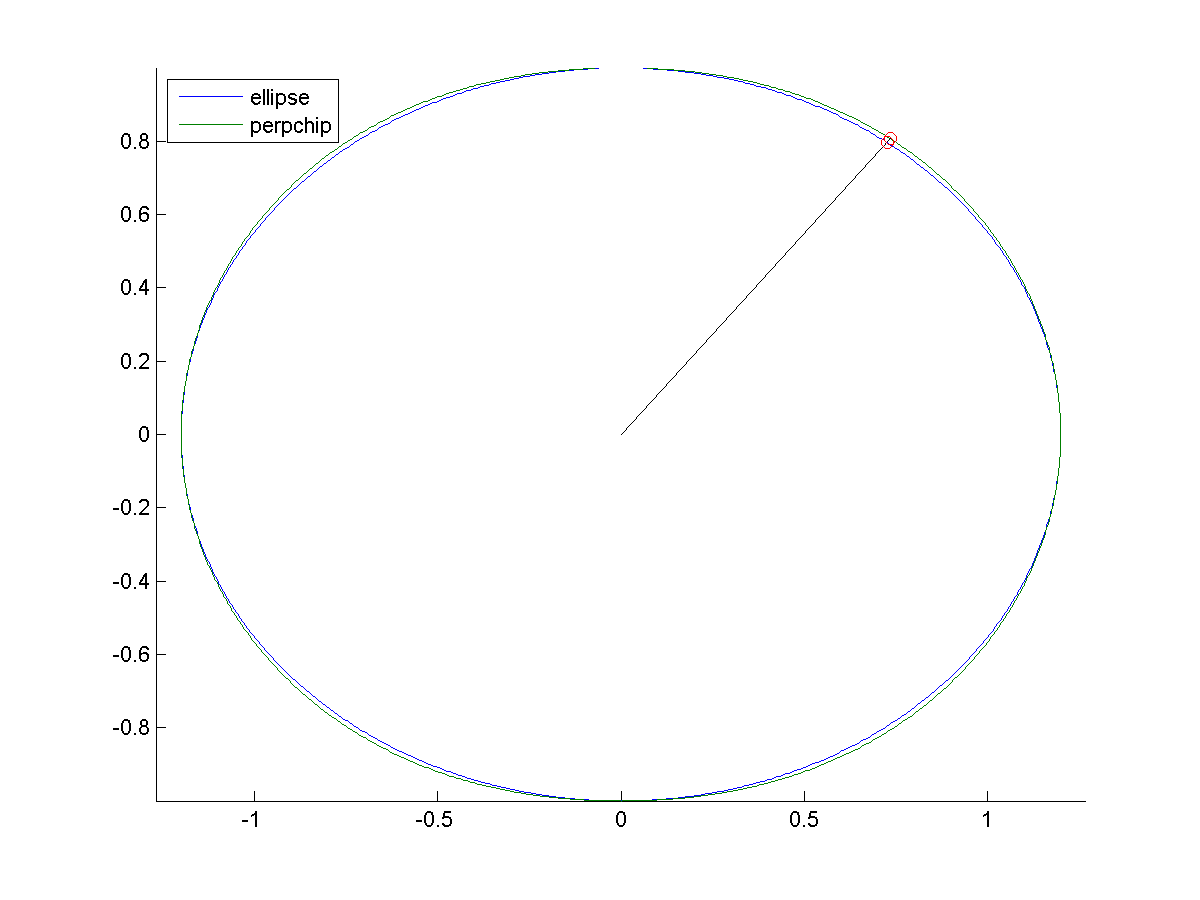}
\end{center} 
\caption{\label{simulation_compare_ellipse1}Comparaison entre l'ellipse et la courbe donnée par \texttt{perpchip}
pour $a=1.2$ et $b=1.0$.}
\end{figure}

Pour $a=1.2$ et $b=1.0$, on a obtenu 
un écart relatif en $r$, $\varepsilon_r=1.27\% $
et en surface égal à $\Delta S=1.24\%$, soit donc petits et il y aura donc peu d'écart
entre les résultats obtenus avc notre méthode et les ellipses de Henderson.
Voir aussi la figure \ref{simulation_compare_ellipse1}.

On a aussi fait varier $a$ dans $[a_{\min},a_{\max}]$ et $b$ dans $[b_{\min},b_{\max}]$
avec respectivement $n_a$ et $n_b$ valeurs. 
Pour les paramètres suivants, 
\begin{align*}
&a_{\min}=0.8,\\
&a_{\max}=1.2,\\
&n_a=30,\\
&b_{\min}=0.8,\\
&b_{\max}=1.2,\\
&n_b=30,
\end{align*}
l'écart maximum obtenu en $r$ vaut
$6.17\% $
et en surface égal à $6.19\%$, soit donc petits.

\begin{figure}[h] 
\begin{center} 
\includegraphics[width=10 cm]{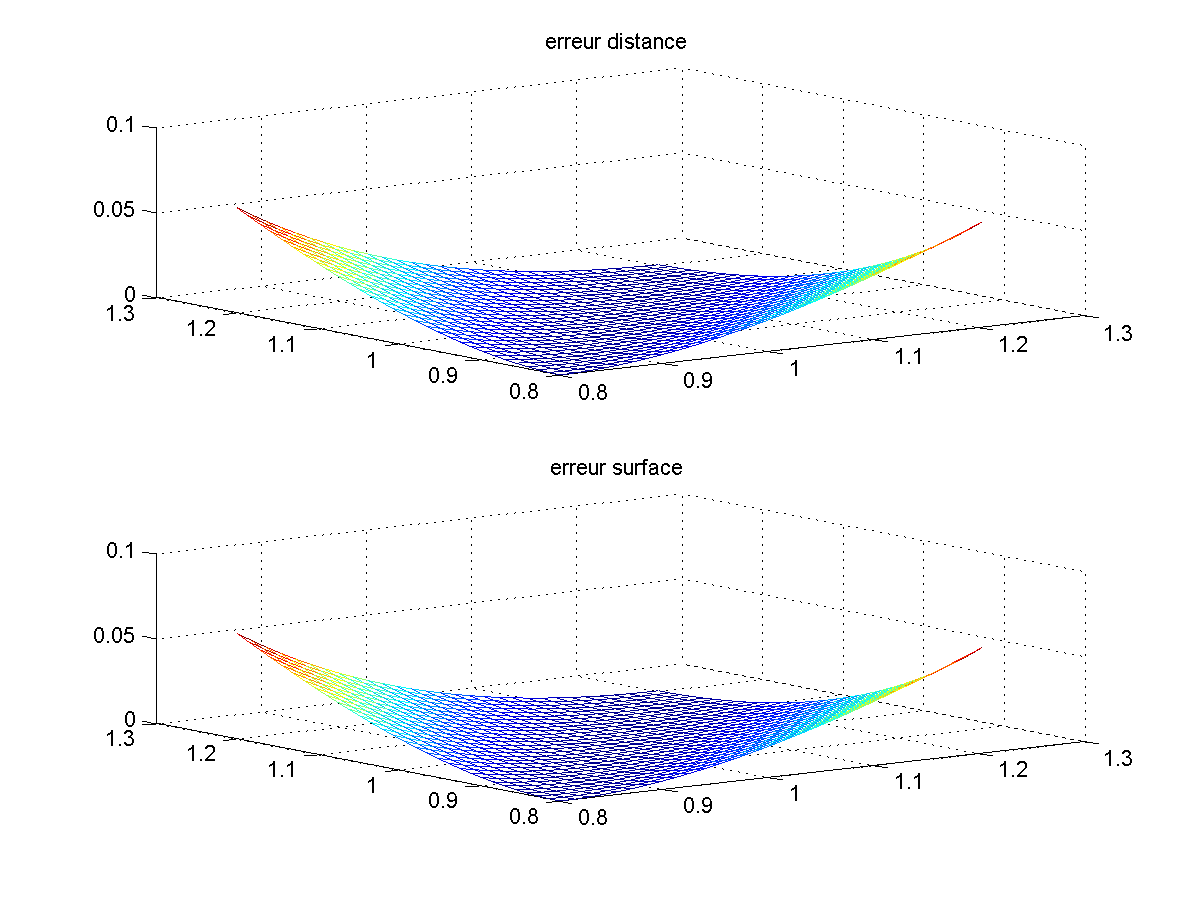}
\end{center} 
\caption{\label{simulation_compare_ellipse2}Comparaison entre l'ellipse et la courbe donnée par \texttt{perpchip}
pour $a$ et $b$ décrivant $[a_{\min},a_{\max}]$ et $[b_{\min},b_{\max}]$.}
\end{figure}

Voir aussi la figure \ref{simulation_compare_ellipse2}.

\begin{figure}
\centering
%%% sous figure 
\subfigure[\label{simulation_compare_ellipse10_1}$b=1.3$ : $\varepsilon_r=2.60\% $
et $\Delta S=2.57\%$]
%{\epsfig{file=simulation_compare_ellipse10_1.eps, width=7cm}}
{\includegraphics[width=7 cm]{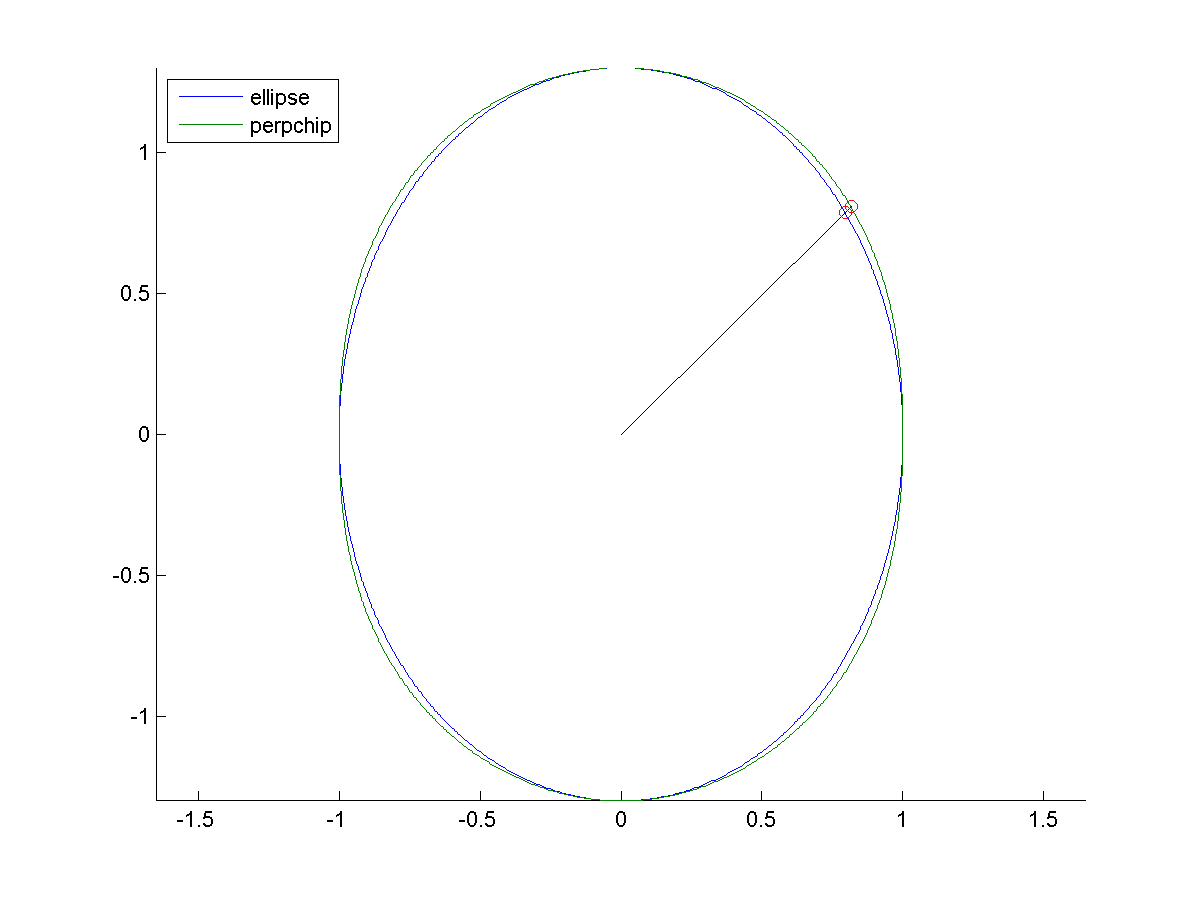}}
%%% sous figure 
\subfigure[\label{simulation_compare_ellipse10_2}$b=1.5$ : $\varepsilon_r=6.17\% $
et $\Delta S=6.19\%$]
%{\epsfig{file=simulation_compare_ellipse10_2.eps, width=7cm}}
{\includegraphics[width=7 cm]{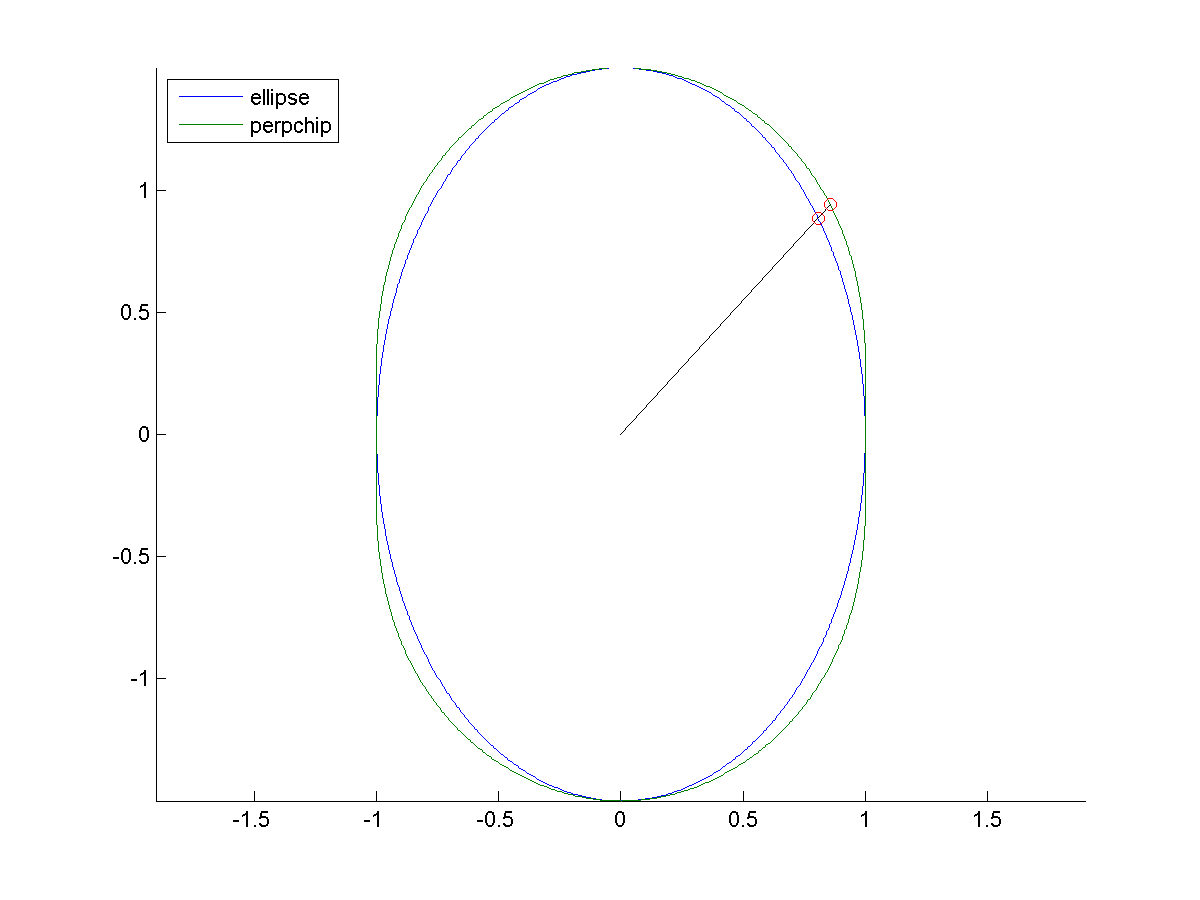}}
%%% sous figure 
\subfigure[\label{simulation_compare_ellipse10_3}$b=2.0$ : $\varepsilon_r=18.10\% $
et $\Delta S=18.57\%$]
%{\epsfig{file=simulation_compare_ellipse10_3.eps, width=7cm}}
{\includegraphics[width=7 cm]{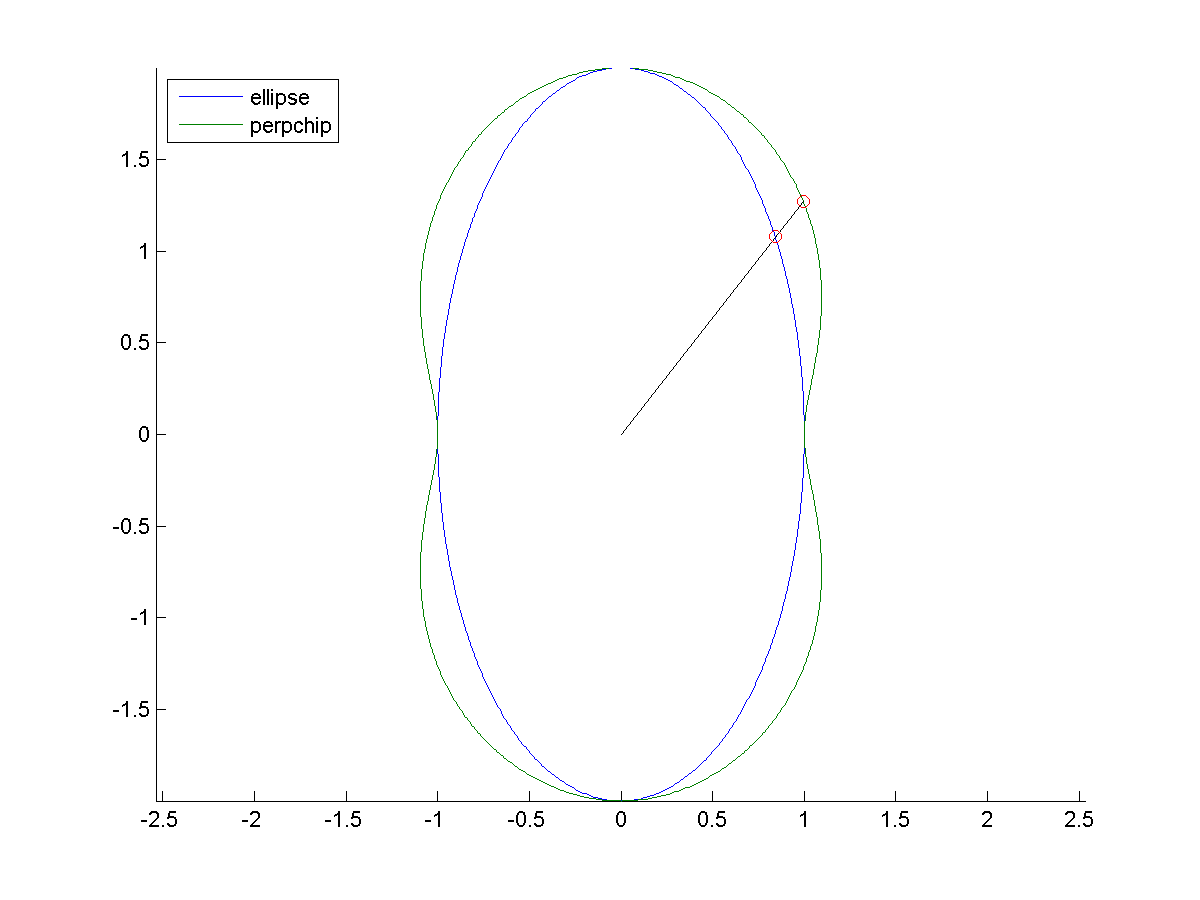}}
%%% sous figure 
\subfigure[\label{simulation_compare_ellipse10_4}$b=4.0$ : $\varepsilon_r=75.19\% $
et $\Delta S=83.57\%$]
%{\epsfig{file=simulation_compare_ellipse10_4.eps, width=7cm}}
{\includegraphics[width=7 cm]{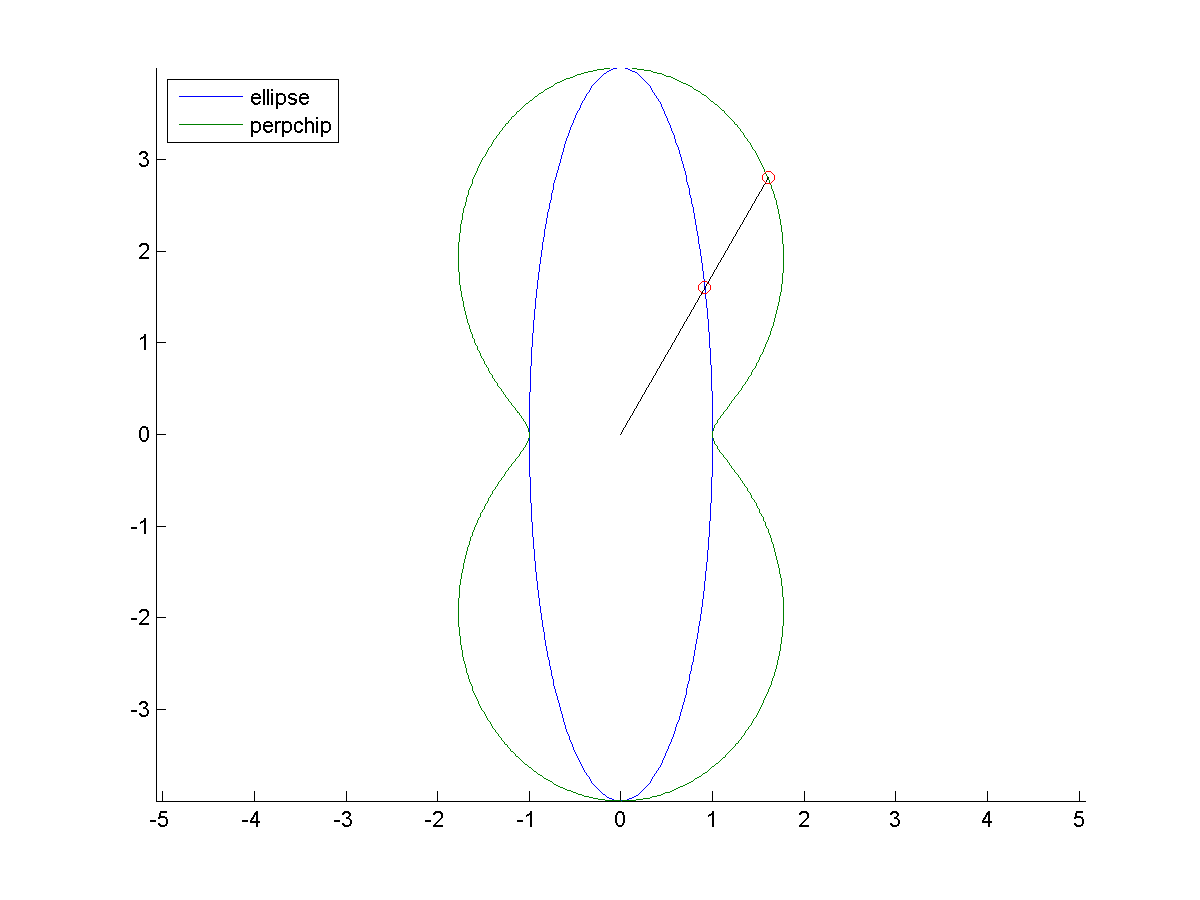}}
\caption{\label{simulation_compare_ellipse3}Comparaison entre l'ellipse et la courbe donnée par \texttt{perpchip} pour $a=1.0$ et $b$ variant.}
\end{figure}

Profitons-en aussi pour tracer des ellipses en faisant varier $b$ :
on choisit
$a=1.0$
et 
$b$ prenant ses valeurs dans $\{%
1.3,
1.5,
2.0,
4.0
\}$.
Voir la figure \ref{simulation_compare_ellipse3}.

%%%%%%%%%%%%%%%%%%%%%%%%%%%%%%%%%%%%%%%%%%%%%%%%%%%%%%%%%%%%%

Dans certains cas, les ellipses de Henderson paraissent donner des meilleurs résultats, notamment quand les 
ellipses sont trop 
dissymétrique (voir par exemple figures \ref{simulation_compare_ellipse10_3}
et  \ref{simulation_compare_ellipse10_4}), où les profils ne semblent pas convexes.
Cela n'est pas contraire à la monotonie de $r$. On peut établir par un critère si des profils ne sont pas convexes.
En effet (on renvoie à \cite[p. 96]{guinin94} ou \cite[p. 24]{bastienmt25}) : une fonction
définie en polaire par $r(\theta)$ est convexe et l'origine appartient à la convavité en au point défini par l'angle $\theta$
si et seulement si 
\begin{equation}
\label{comelleq51}
r^2(\theta)+2r'^2(\theta)-r(\theta) r''(\theta)>0,
\end{equation}
ce que l'ont sait établir si $r$ est un polynôme de degré 3.

%%%%%%%%%%%%%%%%%%%%%%%%%%%%%%%%%%%%%%%%%%%%%%%%%%%%%%%%%%%%%
%\input{simulation_validation_ellipsoide_convexite}
% fichier tex crée par MaTeXBuild02 le 23-Jun-2014 11:49:25
% à compiler avec MaTeXBuild02('simulation_validation_ellipsoide_convexite',0)
% après avoir fait  clear all;MaTeXBuild02('validation_ellipsoide',0)

Par exemple, 
% vérifier bien cela si autres simulations !!!
\ifcase 0
% cas =0 
aucun des profils des figures \ref{valelifig10} n'est convexe, ce qui est surtout visible pour les petites valeurs de $N$.
\or
% cas =1
il existe des profils des figures \ref{valelifig10} non convexes.
\fi

%%%%%%%%%%%%%%%%%%%%%%%%%%%%%%%%%%%%%%%%%%%%%%%%%%%%%%%%%%%%%

%%%%%%%%%%%%%%%%%%%%%%%%%%%%%%%%%%%%%%%%%%%%%%%%%%%%%%%%%%%%%
%\input{verifie_convexite_tot}
% fichier tex crée par MaTeXBuild02 le 23-Jun-2014 11:53:01
% à compiler avec MaTeXBuild02('verifie_convexite_tot',0)
% après avoir fait clear all;MaTeXBuild02('exemple_lezard',0)

\ifcase 1
% cas =0 
Aucun des profils des figures \ref{exemple_lezard_queue} n'est convexe.
\or
% cas =1
\ifcase 0
Pour la figure  \ref{exemple_lezard_queue}, seul le profil d'indice $5$ n'est pas convexe.
\or
Il existe des profils des figures \ref{exemple_lezard_queue} non convexes, dont les valeurs des numéros sont l'ensemble : 
${\5\}$.
\fi
\fi

\ifcase 1
% cas =0 
Aucun des profils des figures \ref{exemple_lezard_tronc} n'est convexe.
\or
% cas =1
\ifcase 0
Pour la figure  \ref{exemple_lezard_tronc}, seul le profil d'indice $16$ n'est pas convexe.
\or
Il existe des profils des figures \ref{exemple_lezard_tronc} non convexes, dont les valeurs des numéros sont l'ensemble : 
${\16\}$.
\fi
\fi

%%%%%%%%%%%%%%%%%%%%%%%%%%%%%%%%%%%%%%%%%%%%%%%%%%%%%%%%%%%%%

En l'absence de convexité, il suffira de rajouter des points de mesure comme le permet notre méthode.

%%%%%%%%%%%%%%%%%%%%%%%%%%%%%%%%%%%%%%%%%%%%%%%%%%%%%%%%%%%%%
%%%%%%%%%%%%%%%%%%%%%%%%%%%%%%%%%%%%%%%%%%%%%%%%%%%%%%%%%%%%%
\section{Conclusion}
\label{conclusion}

Comme observé dans la section \ref{simulation}, les 4 méthodes donnent des résultats très semblables.
La méthode A nous semble donc à privilégier dans la mesure où, découplée par plan de coupe, elle 
permet de réduire les calculs aux petits nombres de données expérimentalement par coupe.

%%%%%%%%%%%%%%%%%%%%%%%%%%%%%%%%%%%%%%%%%%%%%%%%%%%%%%%%%%%%%
%%%%%%%%%%%%%%%%%%%%%%%%%%%%%%%%%%%%%%%%%%%%%%%%%%%%%%%%%%%%%
\appendix
%%%%%%%%%%%%%%%%%%%%%%%%%%%%%%%%%%%%%%%%%%%%%%%%%%%%%%%%%%%%%
%%%%%%%%%%%%%%%%%%%%%%%%%%%%%%%%%%%%%%%%%%%%%%%%%%%%%%%%%%%%%
\section{Interpolation polynomiale d'hermite cubique  périodique par morceaux : les fonctions \texttt{perpchip} et \texttt{perspline}}
\label{46580percubint_Jbastien_2014}

\selectlanguage{english}

For more details, the reader is referred to \cite{46580percubint_Jbastien_2014}\footnote{\citefield{46580percubint_Jbastien_2014}{note}}.

\begin{figure}[h] 
% \psfrag{x0}{$x_{1}$}
% \psfrag{x1}{$x_{2}$}
% \psfrag{x2}{$x_{3}$}
% \psfrag{xi}{$x_{i}$}
% \psfrag{xNmu}{$x_{N-1}$}
% \psfrag{xN}{$x_{N}$}
% \psfrag{yi}{$y_{i}$}
% \psfrag{y1}[][l]{$y_{1}=y_N$}
% \psfrag{perp}{\texttt{perpchip}}
% \psfrag{persp}{\texttt{perspline}}
\begin{center} 
\includegraphics[width=11 cm]{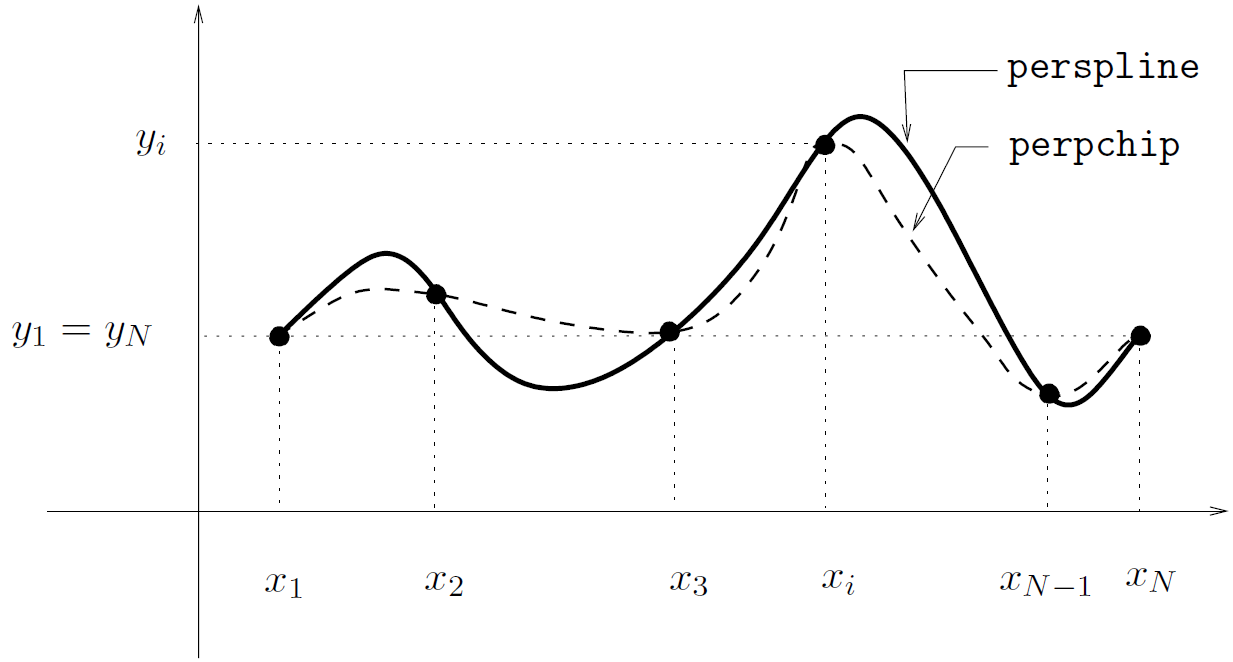}
\end{center} 
\caption{\label{fig101}The principle of periodical interpolation.} 
\end{figure}

Let $N\in \En^*$ with $n\geq 2$ and ${(x_i)}_{1\leq i \leq N}$,  ${(y_i)}_{1\leq i \leq N}$
be $2N$ reals numbers. We assume without  loss of generality that 
\begin{equation}
\label{exeq01}
x_1<x_2<\hdots<x_N.
\end{equation}
We try to find a function $f$ such that 
\begin{equation}
\label{exeq10}
\forall k\in \{1,\hdots,N\},\quad
f(x_k)=y_k.
\end{equation}
This interpolation problem is very classical and there are a lot of solutions.
For example, cubic piecewise interpolation are provided by the following matlab functions : 
\begin{itemize}
\item[$\bullet$] \path|spline|, written by  Carl de Boor ;
see \cite{MR1900298} or \cite{MR2065657} ; 
\item[$\bullet$] \path|pchip| ( so called for <<~Piecewise Cubic Hermite Interpolating Polynomial~>>) ;
see 
\cite[chap. 3]{MR567271}, \cite[p. 104]{kahaner1988} or \cite{MR2065657}.
\end{itemize}
These both functions ensure that $f$ is of class $C^1$ on the interval $[x_1,x_N]$.
Moreover, (see the help of function \path|pchip|)
the functions supplied by   \path|spline| and  \path|pchip|  are  constructed in exactly the same way,
    except that the slopes at the $x_i$ are chosen differently. This has the following effects: 
    \path|spline| is smoother, i.e., $f''$ is continuous.
    \path|spline| is more accurate if the data are values of a smooth function.
    \path|pchip| has no overshoots and less oscillation if the data are not smooth.
    \path|pchip| is less expensive to set up.
Moreover, we assume that now
\begin{equation}
\label{exeq20}
y_1=y_N.
\end{equation}
Thus, we try to determine $f$ which is $T$-periodical  (with $T=x_N-x_1$). 
This property implies that 
\begin{equation}
\label{exeq30}
f'(x_1+0)=f'(x_N-0).
\end{equation}
The functions \path|pchip| and  \path|spline|
of matlab are adapted to the periodical case:
\path|perpchip| and  \path|perspline|.

%%%%%%%%%%%%%%%%%%%%%%%%%%%%%%%%%%%%%%%%%%%%%%%%%%%%%%%%%%%%%
%\input{example_perspline_perpchip}
% fichier tex crée par MaTeXBuild02 le 23-Jun-2014 11:25:51
% à compiler avec MaTeXBuild02('example_perspline_perpchip',0)

\begin{figure}[h] 
\begin{center} 
\includegraphics[width=11 cm]{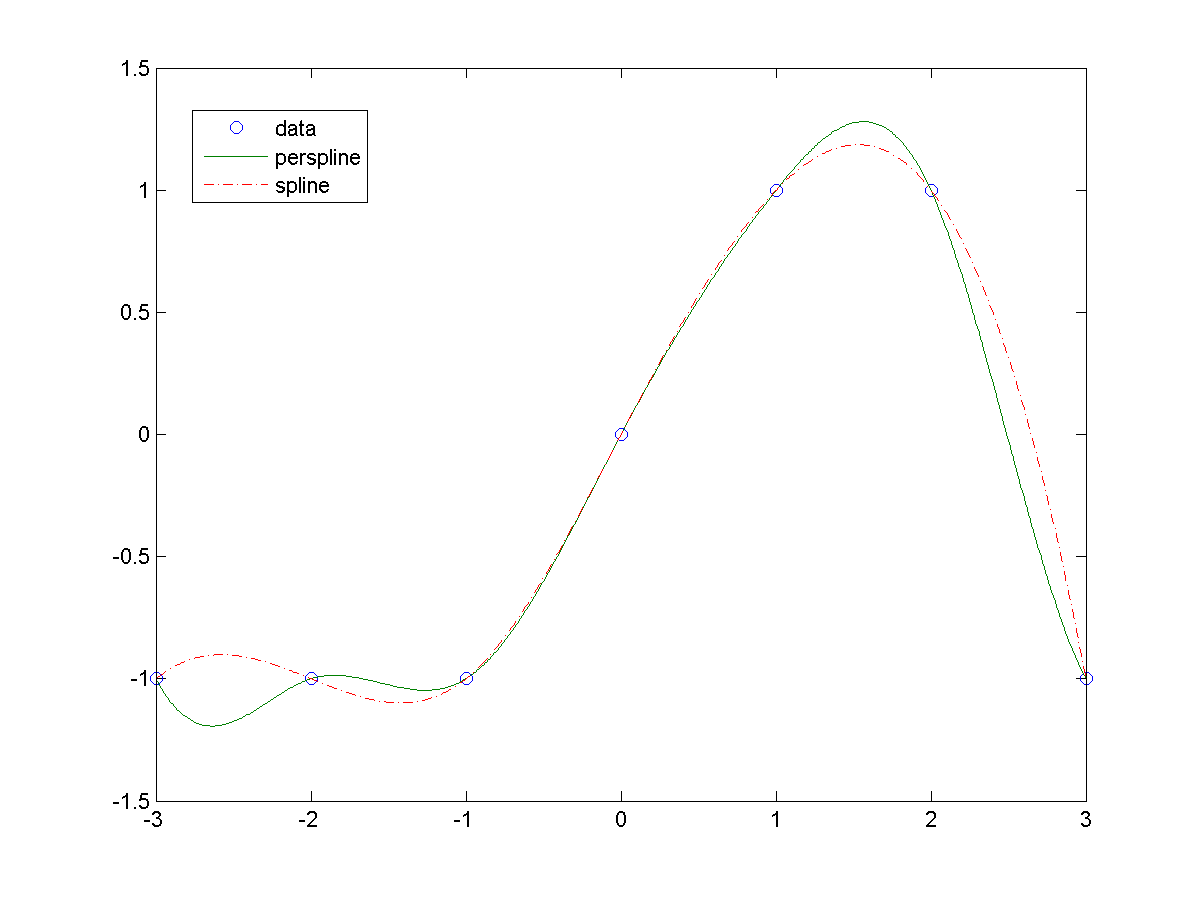}
\end{center} 
\caption{\label{figdudu01}Example for function \texttt{perspline}.} 
\end{figure}

\begin{figure}[h] 
\begin{center} 
\includegraphics[width=11 cm]{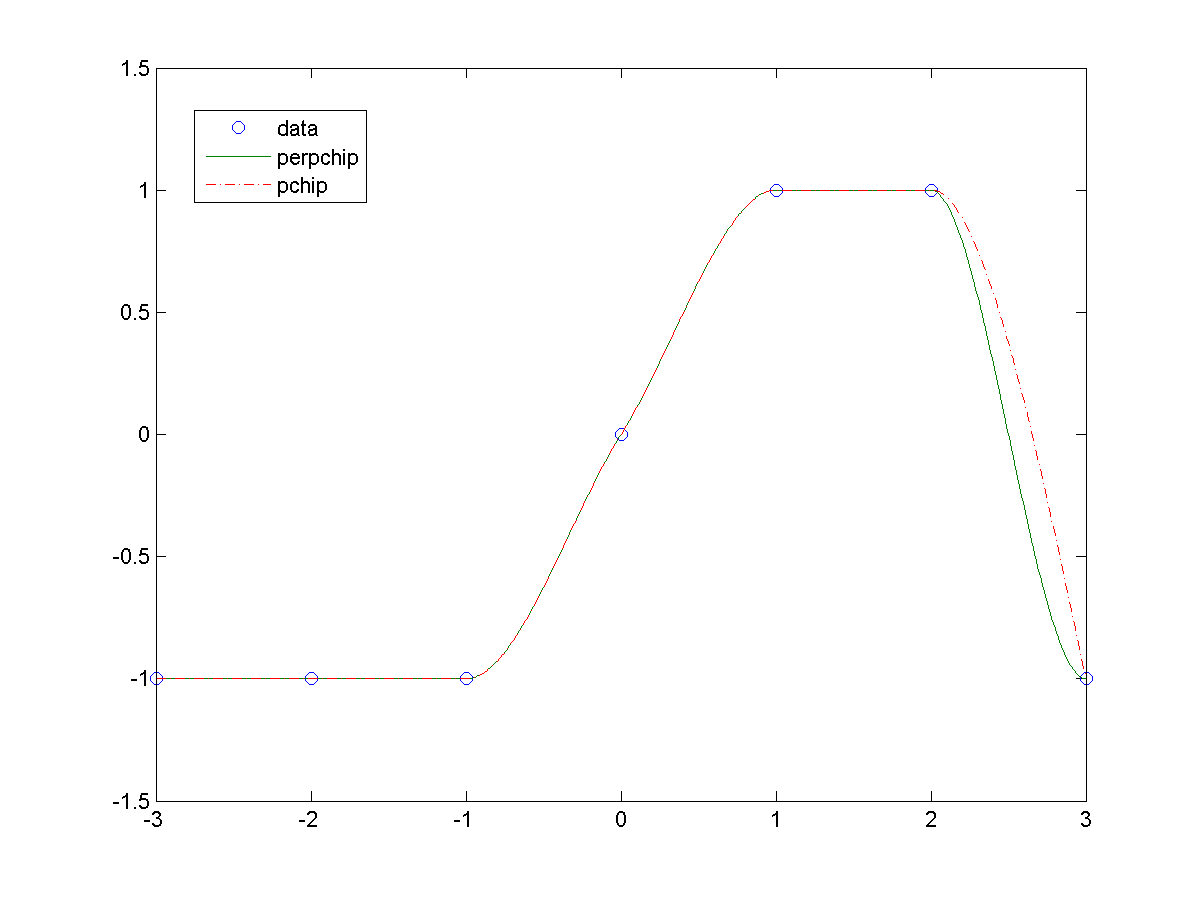}
\end{center} 
\caption{\label{figdudu02}Example for function \texttt{perpchip}.} 
\end{figure}

See for example figures \ref{figdudu01} and  \ref{figdudu02}.

%%%%%%%%%%%%%%%%%%%%%%%%%%%%%%%%%%%%%%%%%%%%%%%%%%%%%%%%%%%%%%

\selectlanguage{french}

%%%%%%%%%%%%%%%%%%%%%%%%%%%%%%%%%%%%%%%%%%%%%%%%%%%%%%%%%%%%%
%%%%%%%%%%%%%%%%%%%%%%%%%%%%%%%%%%%%%%%%%%%%%%%%%%%%%%%%%%%%%
\section{Développement du polynôme $P$ en fonction des valeurs particulière de $A$ et de $B$}
\label{calcul_poly} 

Le polynôme $P$ est donné par \eqref{eq223} et \eqref{eq224}.
On peut utiliser la formule du binôme de newton pour donner la forme générale des coefficients 
$c_i$ et des exposants $\alpha_i$ et $\beta_i$, mais seules un nombre finis de valeurs de $A$ et de $B$ seront utilisés.
On vérifie donc à la main que 
\begin{align*}
&A=0\text{ et } B=0
\Longrightarrow
P(X,Y)=1,\\
&A=1\text{ et } B=0
\Longrightarrow
P(X,Y)=a+X,\\
&A=0\text{ et } B=1
\Longrightarrow
P(X,Y)=b+Y,\\
&A=2\text{ et } B=0
\Longrightarrow
P(X,Y)=a^2+2aX+X^2,\\
&A=0\text{ et } B=2
\Longrightarrow
P(X,Y)=b^2+2bY+Y^2,\\
&A=1\text{ et } B=1
\Longrightarrow
P(X,Y)=ab+bX+aY+XY.
\end{align*}
On en déduit donc les valeurs de $N$, 
${(c_i)}_{0\leq i \leq N} $,
${(\alpha_i)}_{0\leq i \leq N} $  et 
${(\beta_i)}_{0\leq i \leq N} $ 
dans le tableau \ref{table01}.

\begin{table}
\begin{center}
\begin{tabular}{|l|l|l|l|l|l|}
\hline
$A$ & $B$ & $N$& ${(c_i)}_{0\leq i \leq N}\in $  & ${(\alpha_i)}_{0\leq i \leq N}$ & ${(\beta_i)}_{0\leq i \leq N}$ \\
\hline
0&0& 
   0 &$(1)$ &$(0)$ &$(0)$  
\\ \hline
1&0& 
   1 &$(a,1)$ &$(0,1)$ &$(0,0)$   
\\ \hline
0&1& 
   1 &$(b,1)$ &$(0,0)$ &$(0,1)$  
\\ \hline
 2&0& 
    2 &$(a^2,2a,1)$ &$(0,1,2)$ &$(0,0,0)$  
 \\ \hline
 0&2& 
    2 &$(b^2,2b,1)$ &$(0,0,0)$ &$(0,1,2)$  
 \\ \hline
 1&1& 
    3 &$(ab,b,a,1)$ &$(0,1,0,1)$ &$(0,0,1,1)$  
 \\ \hline
\end{tabular}
\end{center}
\vspace{0.7 cm}
\caption{\label{table01}Valeurs de $N$, 
${(c_i)}_{0\leq i \leq N} $,
${(\alpha_i)}_{0\leq i \leq N} $  et 
${(\beta_i)}_{0\leq i \leq N} $ en fonction
de $A$, $B$, $a$ et $b$.}
\end{table}

%%%%%%%%%%%%%%%%%%%%%%%%%%%%%%%%%%%%%%%%%%%%%%%%%%%%%%%%%%%%%
%%%%%%%%%%%%%%%%%%%%%%%%%%%%%%%%%%%%%%%%%%%%%%%%%%%%%%%%%%%%%
% Bibliograhie  faite avec biblatex
\printbibliography

\end{document}